\documentclass[aps, prb, superscriptaddress,amsmath,amssymb,twocolumn, longbibliography]{revtex4-2}
\usepackage{graphicx}
\usepackage{subfigure}
\usepackage{adjustbox}
\usepackage{bm}
\usepackage{color}
\usepackage{braket}
\usepackage{standalone}
\usepackage{multirow}
\usepackage{tikz}
\usepackage{mathrsfs}
\usepackage{dsfont}
\usepackage{comment}
\usepackage{amsmath}
\usepackage{amsmath,blkarray,booktabs}
\usepackage{float}
\usepackage{enumitem}
\usepackage[colorlinks,bookmarks=true,citecolor=blue,linkcolor=red,urlcolor=blue]{hyperref}
\usepackage{cleveref}
\usepackage{orcidlink}
\usepackage{appendix}
\usepackage{graphicx}
\usepackage{caption}
\usepackage{xcolor}

\newcommand{\bq}{{\mathbf q}}

\newcommand{\br}{{\mathbf r}}

\newcommand{\bs}{{\mathbf s}}

\newcommand{\bp}{{\mathbf p}}

\newcommand{\bv}{{\mathbf v}}

\newcommand{\bK}{{\mathbf K}}
\newcommand{\bM}{{\mathbf M}}

\newcommand{\btau}{\bm{\tau}}

\newcommand{\hc}{\hat c}
\newcommand{\hd}{\hat d}

\newcommand{\hcd}{\hat c^\dagger}
\newcommand{\hdd}{\hat d^\dagger}
\newcommand{\hpsi}{\hat \psi}
\newcommand{\hpsid}{\hat \psi^\dagger}

\newcommand{\cV}{{\mathcal V}}

\newcommand{\cE}{{\mathcal E}}
\newcommand{\cK}{{\mathcal K}}
\newcommand{\bea}{\begin{eqnarray*}}
\newcommand{\eea}{\end{eqnarray*}}
\newcommand{\bean}{\begin{eqnarray}}
\newcommand{\eean}{\end{eqnarray}}
\newcommand{\lgl}{\langle}
\newcommand{\rgl}{\rangle}
\newcommand{\nn}{\nonumber\\}

\newcommand{\dwa}{\downarrow}
\newcommand{\uzf}{{u_{z,F}}}
\newcommand{\uzh}{{u_{z,H}}}
\newcommand{\unf}{{u_{\perp,F}}}
\newcommand{\unh}{{u_{\perp,H}}}

\definecolor{BOcolor}{RGB}{0, 128, 0}
\definecolor{CBOcolor}{RGB}{0, 255, 255}
\definecolor{SVExcolor}{RGB}{210, 180, 140}
\definecolor{SVEcolor}{RGB}{152, 251, 152}
\definecolor{coexistcolorz}{RGB}{255, 105, 180}
\definecolor{coexistcolorv}{RGB}{255, 0, 255}
\definecolor{coexistcolorzv}{RGB}{255, 165, 0}
\newcommand{\FMcircle}{\tikz{\draw[draw=black, fill=white] (0,0) circle (0.75ex)}}
\newcommand{\CDWcircle}{\tikz{\draw[draw=black, fill=red] (0,0) circle (0.75ex)}}
\newcommand{\BOcircle}{\tikz{\draw[draw=black, fill=BOcolor] (0,0) circle (0.75ex)}}
\newcommand{\CBOcircle}{\tikz{\draw[draw=black, fill=CBOcolor] (0,0) circle (0.75ex)}}
\newcommand{\AFcircle}{\tikz{\draw[draw=black, fill=yellow] (0,0) circle (0.75ex)}}
\newcommand{\CAFcircle}{\tikz{\draw[draw=black, fill=blue] (0,0) circle (0.75ex)}}
\newcommand{\SVEcircle}{\tikz{\draw[draw=black, fill=SVEcolor] (0,0) circle (0.75ex)}}
\newcommand{\SVExcircle}{\tikz{\draw[draw=black, fill=SVExcolor] (0,0) circle (0.75ex)}}
\newcommand{\BCcircle}{\tikz{\draw[draw=black, fill=coexistcolorz] (0,0) circle (0.75ex)}}
\newcommand{\CAcircle}{\tikz{\draw[draw=black, fill=coexistcolorv] (0,0) circle (0.75ex)}}
\newcommand{\CCcircle}{\tikz{\draw[draw=black, fill=coexistcolorzv] (0,0) circle (0.75ex)}}

\begin{document}

\title{Uniquely identifying quantum Hall phases in charge neutral graphene}
\author{Jincheng An}
\email{Jincheng.An@uky.edu}
\affiliation{Department of Physics and Astronomy, University of Kentucky, Lexington, KY 40506, USA}

\author{Ganpathy Murthy\orcidlink{0000-0001-8047-6241}}
\email{murthy@g.uky.edu}
\affiliation{Department of Physics and Astronomy, University of Kentucky, Lexington, KY 40506, USA}	
\

\begin{abstract}
Charge-neutral graphene in the quantum Hall regime is an example of a quantum Hall ferromagnet in a complex spin-valley space. This system exhibits a plethora of phases, with the particular spin-valley order parameters chosen by the system depending sensitively on the short-range anisotropic couplings, the Zeeman field, and the sublattice symmetry breaking field. A subset of order parameters related to lattice symmetry-breaking have been observed by scanning tunneling microscopy. However, other order parameters, particularly those which superpose spin and valley, are more elusive, making it difficult to pin down the nature of the phase. We propose a solution this problem by examining two types of experimentally measurable quantities; transport gaps and collective mode dispersions. We find that the variation of the transport gap with the Zeeman and sublattice symmetry breaking fields, in conjunction with the number of Larmor and gapless modes, provides a unique signature for each theoretically possible phase. 

\end{abstract}
\maketitle

\section{Introduction}
\label{sec:Intro}
Since the discovery of the integer and fractional Quantum Hall Effects (QHE)~\cite{IQHE_Discovery_1980,FQHE_Discovery_1982,Laughlin_1983}, the study of two-dimensional (2D) systems under strong perpendicular magnetic fields has become an important subfield of condensed matter physics. Quantum Hall states are known to be the simplest examples of topological insulators. The bulk is gapped to all charged excitations. Consequently, all electric conduction occurs via the edges, which host chiral edge modes. In 2004,  graphene~\cite{Berger_etal_2004,Novoselov_etal_2004,Zhang_etal_2005,neto2009:rmp}, a single layer of carbon atoms arranged in a honeycomb lattice consisting of two sites (A and B) in its unit cell, was discovered. Graphene is the prototype of a large class of 2D materials such as transition metal dichalchogenides, van der Waals heterostructures~\cite{Geim2013VanDW,Novoselov20162DMA}, etc, and has exhibited unique properties including high electron mobility, linear band dispersion, and tunable electronic characteristics. In the past two decades, graphene has become an excellent platform for exploring quantum Hall physics and advancing our understanding of topological states of matter and spontaneous symmetry breaking.\\

Near charge neutrality, the low-energy electrons in graphene reside in two valleys located at the two inequivalent corners $K$ and $K^\prime$ of the first Brillouin zone, related by time-reversal symmetry~\cite{neto2009:rmp}. In each valley, they exhibit a linear dispersion relation, analogous to that of massless Dirac fermions. Upon the application of a perpendicular magnetic field $B$, the linear Dirac spectrum will develop particle-hole symmetric Landau levels $n=0,\ \pm 1,\ \pm 2...$ with energy $E_{\pm n}\propto\sqrt{B|n|}$. In the lowest Landau level, indexed by 
$n=0$, the states in each valley are localized on a specific sublattice, a phenomenon known as sublattice-valley locking~\cite{neto2009:rmp}.\\

\begin{figure}[H]
    \centering \includegraphics[width=0.48\textwidth,height=0.33\textwidth]{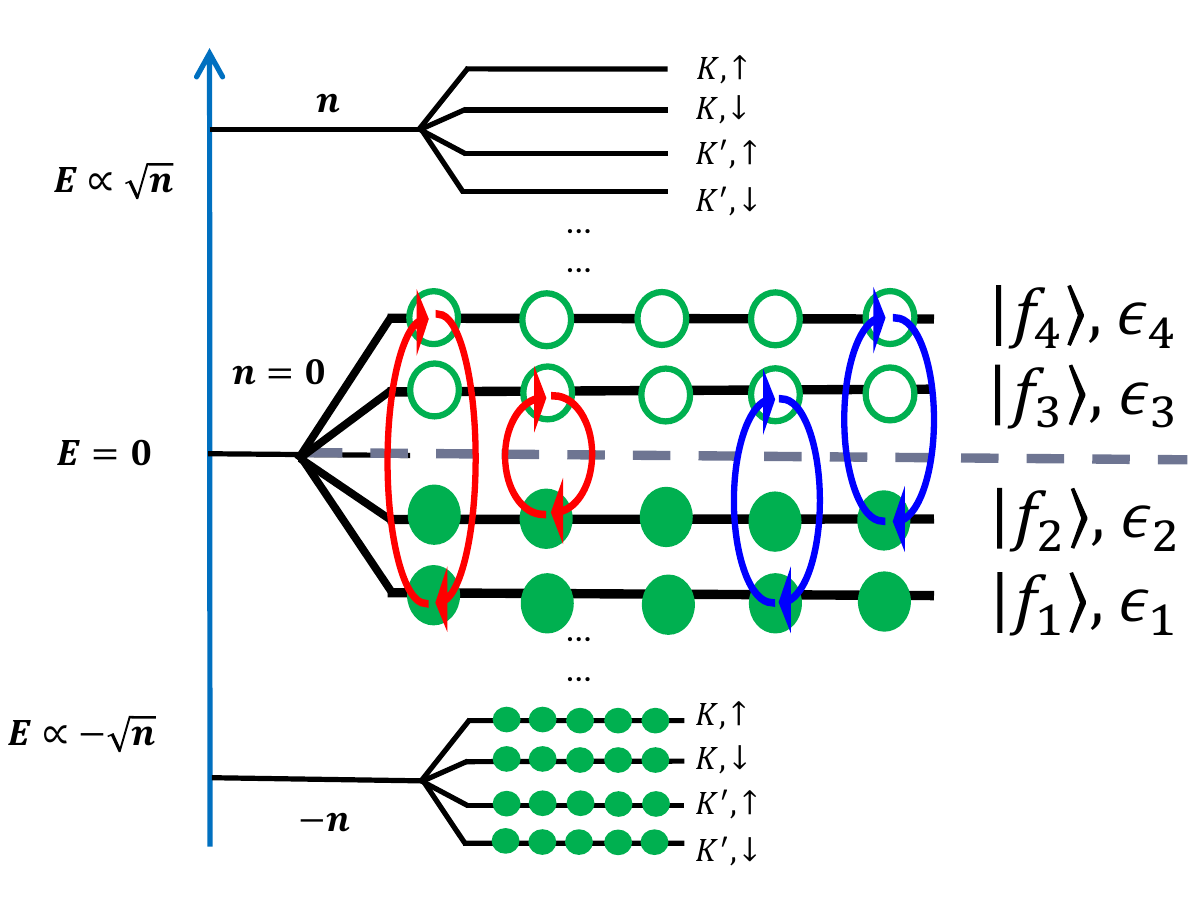}
    \caption{Landau levels in monolayer graphene. At charge neutrality ($\nu=0$), two of the four LLs in the $n=0$ manifold are occupied, while the $n<0$ ($n>0$) LL manifolds are completely occupied (empty). The black dashed line indicates the location of the chemical potential. $|f_i\rgl$'s are the spinors optimized by Hartree Fock approximation and $\epsilon_i$'s are the corresponding eigenvalues. Curved arrows (red and blue, with their difference being explained later) label all possible particle-hole/hole-particle excitations within $n=0$ LL, which form the basis states for collective eigenmodes. Thus, each ground state has four collective modes. }
    \label{graphene_LLs}
\end{figure}

Each Landau level manifold is (almost) fourfold degenerate (see Fig. \ref{graphene_LLs}), which is a combination of twofold valley pseudospin and the twofold electron spin (near) degeneracy. It should be noted that the Zeeman energy $E_Z$ is typically the smallest scale in the problem. In the non-interacting limit, when $E_Z\to 0$, the Hamiltonian is invariant under $SU(4)$ transformations in spin-valley space. The Coulomb interaction between electrons respects the $SU(4)$ symmetry, and  is the dominant energy scale. When a  Landau level (LL) manifold is partly full, the Coulomb interaction spontaneously breaks the $SU(4)$ symmetry by aligning the electrons in specific linear combinations of spin and valley indices. This symmetry-breaking is known as Quantum Hall ferromagnetism (QHFM)~\cite{Shivaji_Skyrmion,QHFM_Yang_etal_1994, QHFM_Moon_etal_1995}. Since the Coulomb interaction is $SU(4)$ symmetric, all orientations in the spin-valley space have the same energy. However, one-body terms that break the $SU(4)$ symmetry, such as the Zeeman coupling and sublattice potential $E_V$ induced by the HBN substrate~\cite{KV_DGG_2013,KV_expt_2013,KV_hBN_Jung2015,KV_hBN_Jung2017},  favor certain spin and valley configurations. For example, at charge neutrality, two of the four available LLs in the $n=0$ manifold must be occupied. If only the Coulomb interaction and the Zeeman coupling are present, then the system will choose to occupy the $|K\uparrow\rangle, |K'\uparrow\rangle$ LLs~\cite{abanin2006:nu0, Brey_Fertig_2006}. This state is predicted to be have a pair of counter-propagating edge modes carrying opposite spin. If one ignores the tiny spin-orbit coupling in graphene~\cite{SOC_Graphene_MacD2006,SOC_Graphene_Sandler2007,SOC_Graphene_Yao2007,SOC_graphene_2010,SOC_graphene_2017}, these modes cannot scatter into each other, resulting in the prediction that at charge neutrality, graphene should be in a quantum spin Hall state~\cite{Kane_Mele_2DTI_PhysRevLett.95.146802} with a two-terminal conductance of $2e^2/h$. \\

Experimentally, it is found that at purely perpendicular field, graphene is not a quantum spin Hall insulator, but rather a vanilla insulator with no edge modes~\cite{young2014:nu0}. Upon increasing $E_Z$ by adding a parallel magnetic field, the two-terminal conductance tends to  $2e^2/h$ at very large $E_Z$. This necessitates a reconsideration of the model Hamiltonian. Based on a large body of previous work~\cite{alicea2006:gqhe, KYang_SU4_Skyrmion_2006, Herbut1, Herbut2,abanin2006:nu0,brey2006:nu0}, Kharitonov constructed a  simple model~\cite{kharitonov2012:nu0} for graphene. Due to lattice symmetries and momentum conservation, at the level of four-Fermi interactions, the number of electrons in each valley is separately conserved, which results in a $U_v(1)$ valley symmetry~\cite{alicea2006:gqhe}. When projected to the $n=0$ LL manifold there are only three possible interaction terms. One is fully $SU(4)$ symmetric, and can be absorbed into the Coulomb interaction. The other two are a valley Ising coupling $u_z$ and a valley $XY$ coupling $u_\perp$. These couplings do not respect the full $SU(4)$ symmetry of the Coulomb interaction, but rather have the reduced symmetry $SU(2)_{s}\otimes U(1)_{v}\otimes Z_2$, where the first factor is the spin rotation symmetry, and the last is the symmetry of exchanging the two sublattices. It should be noted that if one allows for, say, three-body interactions, then the $U(1)_v$ symmetry should be broken down to $Z_3$, because an umklapp term conserving momentum modulo a reciprocal lattice vector can take three electrons from the $K$ valley and put them in the $K'$ valley. Since these residual anisotropic interactions arise at the lattice scale $a$, which is much smaller than the magnetic length $\ell=\sqrt{h/eB}$, Kharitonov made the natural assumption that they can be treated as ultra-short-range (USR) Dirac delta function-like interactions~\cite{kharitonov2012:nu0}. Four phases are found in this model~\cite{kharitonov2012:nu0} using the Hartree-Fock approximation: (i) A ferromagnetic (FM) phase, identical to the quantum spin Hall phase predicted earlier. (ii) A canted antiferromagnetic (CAFM) phase where each sublattice (valley) has a spin canted along the direction of the total field. (iii) A bond-ordered (BO) phase which breaks the lattice symmetry but is a spin singlet. (iv) A charge density wave (CDW) phase where a single sublattice (valley) is occupied by both spins. These four phases break either spin rotation symmetry, or lattice symmetry, but not both.\\ 

Based on a symmetry classification of all ultra-short-range interactions in graphene~\cite{RG1_Graphene_Aleiner_2007,kharitonov2012:nu0,Stefanidis_Sodemann2022}, and after including the effects of electron-phonon coupling, it is estimated that at realistic fields both $u_z$, $u_\perp$ are a few $meV$, with $u_z>0, u_\perp<0$~\cite{RG1_Graphene_Aleiner_2007,Wei_Xu_Sodemann_Huang_LLM_SU4_breaking_MLG_2024}. At purely perpendicular fields, if $u_z>|u_\perp|$ the system is in the CAFM phase. Upon increasing $E_Z$, the system makes a second-order transition into the FM phase. This is consistent with the two-terminal transport measurements. Additional support for this explanation was provided by magnon transmission experiments~\cite{Magnon_transport_Yacoby_2018, Spin_transport_Lau2018,Magnon_Transport_Assouline_2021, Magnon_transport_Zhou_2022}. Subsequently, samples encapsulated on one side were studied by scanning tunneling microscopy~\cite{li2019:stm, STM_Yazdani2021visualizing, STM_Coissard_2022, Yazdani_FQH_STM2023_1, Yazdani_FQH_STM2023_2}, and it was found that such sample break lattice symmetries by forming either a CDW (induced by the hexagonal Boron Nitride (HBN) substrate) or a BO state. Very recently, it has become clear that Landau level mixing \cite{Murthy_Shankar_LLmix,Bishara_Nayak_LLmix,Sodemann_MacDonald_LLmix,RG_Peterson_Nayak_2013,RG_Peterson_Nayak_2014,Wei_Xu_Sodemann_Huang_LLM_SU4_breaking_MLG_2024,Xu2024} cannot be ignored, because it leads to substantial renormalizations of the short-range couplings, and in fact makes them finite range rather than USR. 

When the USR limit is relaxed theoretically\cite{Das_Kaul_Murthy_2022,De_etal_Murthy2022,Stefanidis_Sodemann2023,Atteia_Goerbig_2021,Hegde_Sodemann2022,Lian_Goerbig_2017}, additional phases emerge, including a coexistence phase that exhibits both CAF and BO orders and several spin-valley entangled (SVE) states where the occupied spinors cannot be expressed as direct product of spin and valley Bloch spinors. A more challenging aspect arises as we move away from charge neutrality, entering the fractional quantum Hall regime, where the FQHE states exhibit strong correlations. The QHFM states in FQHE were first studied in \cite{Sodemann_MacDonald_2014,Hegde_Sodemann2022} under the assumption of USR interactions, where it was found that at $\nu=-1/3$, for example, in addition to the four phases found by Kharitonov~\cite{kharitonov2012:nu0} at $\nu=0$, an additional antiferromagnetic phase is found. Relaxing the USR assumption for FQHE states leads to a plethora of phases~\cite{Jincheng2024,an2024fractional} exhibiting simultaneous spontaneous symmetry-breaking of both lattice and magnetic symmetries.\\ 



Given this profusion of theoretically possible phases\cite{Das_Kaul_Murthy_2020,Das_Kaul_Murthy_2022,Stefanidis_Sodemann2023,Atteia_Goerbig_2021,Hegde_Sodemann2022,Lian_Goerbig_2017,Jincheng2024,an2024fractional}, it is important to ask how one is to distinguish them experimentally. One longstanding experimental technique is transport~\cite{Zhang2005ExperimentalOO,Young2012SpinAV}, which can measure the charge gap in the bulk. A second set of observables are the collective modes~\cite{Halperin1984} and their dispersions. In $GaAs$ heterostructures the collective mode dispersions have been mapped out for all $q$~\cite{Pinczuk_1993,vKlitzing2009,Pinczuk2024,Liang2024EvidenceFC}. To the best of our knowledge, no such body of work exists yet for graphene. As shown in Fig.~\ref{graphene_LLs}, at $\nu=0$, there are four collective modes, representing linear combinations of particle-hole pairs between the empty and full Landau levels. 

\begin{figure}[H]
    \centering \includegraphics[width=0.44\textwidth,height=0.28\textwidth]{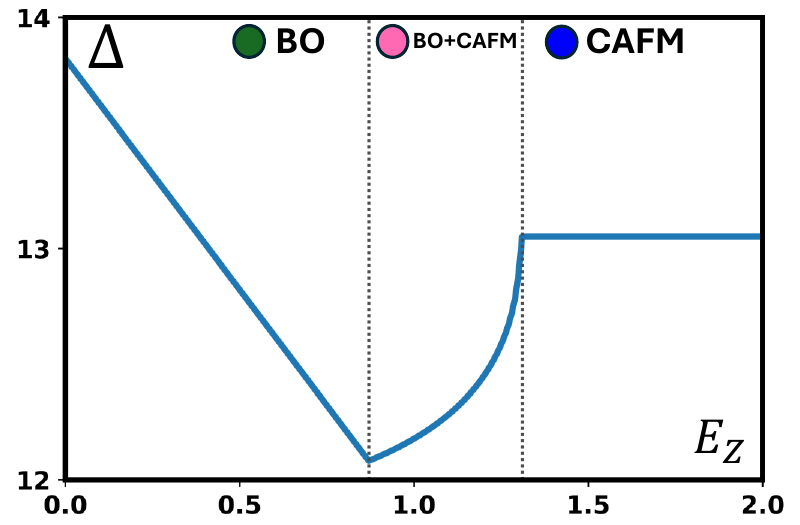}
    \caption{Response of the transport gap $\Delta$ to Zeeman energy $E_Z$ with vanishing valley Zeeman energy. As $E_Z$ increases, a BO$\rightarrow$BO+CAFM$\rightarrow$CAFM phase transition will happen, during which the gap will decrease first then increase and finally level out.}
    \label{coex_gap}
\end{figure}

One of our criteria for uniquely distinguishing phases is the presence of lack thereof of  gapless modes in a given bulk phase. Such modes can be detected, in principle, by nonlocal transport~\cite{JunZhu_2021}, or heat transport~\cite{Parmentier_2024_Heat_Flow,Anindya_Das_2024_Heat_transport}. A more recent technique in the context of graphene is STM \cite{li2019:stm, STM_Yazdani2021visualizing, STM_Coissard_2022, Yazdani_FQH_STM2023_1, Yazdani_FQH_STM2023_2}, which can detect lattice symmetry breaking at the atomic scale. This will also prove useful in distinguishing phases. 
\begin{widetext} 

\begin{table}[h]
\centering
\begin{tabular}{|p{1.2cm}|p{1.2cm}|p{1.9cm}|p{1.7cm}|p{1.2cm}|p{1.2cm}|p{1.2cm}|p{2.1cm}|p{2.1cm}|p{1.7cm}|}
        \hline
         Phase & \FMcircle\  FM &\AFcircle\ \CAFcircle\ (C)AFM & \BOcircle\ \CBOcircle\ (C)BO &\CDWcircle\ CDW &\SVExcircle\ SVEX &\SVEcircle\ SVE & \BCcircle\ BO+CAFM & \CAcircle\ CBO+AFM & \CCcircle COEX\\
        \hline
    $\quad\frac{d\Delta}{dE_Z}$  & $\quad +$ & $\quad 0$ & $\quad -$  & $\quad -$  & $\quad 0$  & $\quad 0$ & \quad +& Undefined & \quad$\pm$\\
        \hline
    $\quad\frac{d\Delta}{dE_V}$  & $\quad -$ & $\quad -$  & $\quad 0$ & $\quad +$ & $\quad 0$ & $\quad 0$ & Undefined& \quad + &  \quad$\pm$\\
        \hline
        $\quad N_G$ & $\quad$0 & $\quad$1 (1) &$\quad$1 (0) & $\quad$0 & $\quad$2 (1) & $\quad$1 & $\quad$2 (1) &\quad 3 (2) &\quad 2 (1)\\
        \hline 
       $\quad N_L$ & $\quad$1 & $\quad$1 & $\quad$0 & $\quad$0 & $\quad$1 & $\quad$1 & $\quad$1 &\quad 0 & \quad 1\\
        \hline 
\end{tabular}
\caption{Unique diagnostics for the possible phases in graphene at $\nu=0$. The first row tells us the response of the transport gap $\Delta$ to increasing $E_Z$. The entry under CBO+AFM is "Undefined" because this phase only exists at $E_Z=0$. The second row tells us the response of $\Delta$ to $E_V$. The entry under BO+CAFM is "Undefined" because this phase only occurs at $E_V=0$. The third row ($N_G$) tells us the number of Goldstone modes. The first number is the nominal number of gapless modes in our model, assuming that $U(1)_\text{valley}$ is an exact symmetry of the Hamiltonian. The number in $()$ is the actual number of Goldstone modes expected when the valley symmetry is reduced to $Z_3$ with the inclusion of 3-body interactions. The last row ($N_L$) tells us whether there is a Larmor mode, indicating a nonzero spin polarization, and the ability of the phase to support magnon transmission. Each column is unique, showing that all the phases can be distinguished from each other. 
}\label{table_1}
\end{table}
\end{widetext}
In this work we will show that, using only the above-mentioned well-established experimental techniques, it is possible to uniquely distinguish all the phases that are theoretically possible at $\nu=0$. The behavior of the transport gaps as the external one-body couplings $E_Z$ and $E_V$ are varied, combined with the presence or absence of gapless collective modes, provides defining diagnostics for each phase. Our central result is Table ~\ref{table_1}, which shows the four quantities which need to be measured in order to distinguish the phases. The signatures are all digital, that is, they can all be represented as the sign of some derivative of the transport gap, or the number of certain collective modes. Each column in Table~\ref{table_1} represents a phase, and the rows represent different physical measurables. As an illustration, let us home in on the most likely phases in real graphene samples at purely perpendicular field and no sublattice symmetry breaking $E_V=0$, which are the the BO, BO+CAF, and CAF phases. $E_V=0$ can be arranged by deliberately misaligning the graphene and the encapsulating HBN layers~\cite{Jung2014OriginOB,hbn_jung_2017,Even_Den_Young2018,RibeiroPalau2018TwistableEW,Finney2019TunableCS}. For certain couplings, these three phases occur in the sequence presented above as $E_Z$ increases, as shown in Fig.~\ref{coex_gap} showing the transport gap as a function of $E_Z$.  The BO phase can be distinguished from the other two by the presence of bond order (STM), and a negative slope of the transport gap with respect to $E_Z$. The BO+CAFM phase is distinguished by bond order, a positive slope of transport gap with respect to $E_Z$ and two special collective modes: A gapless canted antiferromagnetic Goldstone mode, and the Larmor mode at $2E_Z$. The CAF state has no bond order, a transport gap that is independent of $E_Z$, and the same two special modes (Larmor and Goldstone)  as the BO+CAF phase.
In fact, if one can ensure $E_V=0$, the variation of the transport gap with $E_Z$ by itself provides a "smoking gun" signature to distinguish the three different phases. In later sections, after we describe the various theoretically possible phases, we will elucidate the other columns in the table. \\

In addition, we have detailed results for collective mode dispersions in all the theoretically predicted phases. Some of the collective mode results for the four phases predicted in the OM are previously known\cite{MSF2014,Wu_Sodemann_Yasufumi_MacDonald_Jolicoeur_2014,MSF2016,Zapata2017,Wei_Huang_2021,De_Rao_Murthy2024}, but the results for the coexistence (COEX) phase and the SVE phases are new.  \\

The BO and CAF phases have been known theoretically since the work of Kharitonov~\cite{kharitonov2012:nu0}. The potential new phase is the coexistence phase~\cite{Das_Kaul_Murthy_2022,De_etal_Murthy2022}, called either BO+CAF if $E_V=0$, or COEX if $E_V\neq0$. In the previous paragraph we looked at the $E_V=0$ case. Let us examine how the COEX  phase might be distinguished from its neighbors when $E_V\neq0$. Unfortunately, when $E_V\neq0$, the variation of the gap with $E_Z$  in this phase can be non-monotonic, and the behavior of the transport gap with $E_Z,~E_V$ does not by itself provide a conclusive signature of the COEX phase. One must then go to the next step of measuring the symmetry-breaking order parameters and the collective modes. The COEX phase has a nonzero bond order which can be detected in STM, and also a nonzero spin polarization.  Let us go into a bit more detail on the collective modes. Nominally the COEX phase (and also its $E_V\to0$ limit, the BO+CAF phase) has two gapless Goldstone modes. They correspond to long-wavelength rotations of the $XY$-like order in the spin and the valley space. However, generically, when higher-body interactions are included, the valley $U(1)$ symmetry will be reduced to $Z_3$. Thus, the valley collective excitations are expected to be gapped in the BO and COEX phases. Additionally, the presence of a nonzero spin polarization means that a Larmor mode is present. Any phase with a Larmor mode can support spin magnons, which can be verified in magnon transmission experiments~\cite{Magnon_transport_Yacoby_2018, Spin_transport_Lau2018, Magnon_Transport_Assouline_2021, Magnon_transport_Zhou_2022}.  Therefore, in the generic case when $E_V\neq0$, the simultaneous presence of (i) a nonzero bond order, (ii) a Goldstone mode associated with the $XY$ like fluctuations of the canted spin order, and (iii) the Larmor mode, will serve to distinguish the COEX phase from its neighbors (BO and CAFM).

This work is organized as follows: In Sec. \ref{sec2}, we introduce the interacting Hamiltonian and employ the Hartree Fock approximation to obtain the variational energy functional. In Sec. \ref{sec3}, set up the time-dependent Hartree-Fock (TDHF) formalism to find the collective modes.  In Sec. \ref{sec4}, we relax the USR limit of the anisotropic interaction. We do this by expressing the interactions in terms of Haldane pseudopotentials $V_m$\cite{Haldane_Pseudopot1983}, where $m$ is the relative angular momentum of the two electrons that are interacting. The pseudopotential for $m=0$ corresponds to the USR interaction. In addition, to go minimally beyond USR interactions, we introduce  nonzero $m=1$ Haldane pseudopotentials and present phase diagrams including many more phases~\cite{Das_Kaul_Murthy_2022,De_etal_Murthy2022,Stefanidis_Sodemann2023} than found in the USR model~\cite{kharitonov2012:nu0}. In Sec. \ref{sec:Gaps_Collective_Excitations}, transport gaps and dispersions of the collective modes for the various phases are presented. We end with conclusions and discussion in Sec. \ref{sec: discussion_summary}.

\section{Model Hamiltonian \& Hartree-Fock Approximation}
\label{sec2}
\subsection{Interacting Hamiltonian in Lowest Landau Level}
We will start with a general interacting Hamiltonian projected to the $n=0$ LL-manifold of graphene. The $U(1)_{valley}\otimes SU(2)_{spin}$ symmetry forces the interaction to take the following form, with two undetermined functions $V_z(r)$ and $V_x(r)=V_y(r)=V_\perp(r)$ 
\bean
\hat H&=&\frac{1}{2}\int d^2\br_1d^2\br_2\nn
&&:\hpsid_\alpha(\br_1) \tau^{a}_{\alpha\lambda}\hpsi_\lambda(\br_1)V_a(\br_1-\br_2)\hpsid_\beta(\br_2)\tau^{a}_{\beta\eta}\hpsi_\eta(\br_2):\nn
\eean
where $\tau^{a}$ are the valley Pauli matrices in the four-dimensional spin/valley space. After Fourier transformation we obtain,
\bean 
V_a(\br_1-\br_2)=\frac{1}{(2\pi)^2}\int d^2\bq e^{-i \bq\cdot(\br_1-\br_2)}v_a(\bq).
\label{eq:vaq_def}\eean 
Next, we expand the creation and destruction operators in terms of the eigenstates of the non-interacting Hamiltonian in the $n=0$ LL-manifold.
\bean
&&\hpsi_\alpha(\br)=\sum_k\phi_k(\br)\hc_{k,\alpha},\nn
&&\phi_k(\br)=\frac{e^{iky}e^{-\frac{(x-k\ell^2)^2}{2\ell^2}}}{\sqrt{L_y\ell \sqrt\pi}},\ k=\frac{2\pi j}{L_y}.
\eean 
Here, we are using Landau gauge and assuming periodic boundary conditions in the $y$-direction. The Hamiltonian can be rewritten as
\bean
\hat H&=&\frac{1}{2}\sum_\bq\sum_{k_1,k_2} \cV_a(\bq,k_1,k_2)\nn
&&:\tau^{a}_{\alpha\lambda}\hcd_{k_1-q_y,\alpha}\hc_{k_1,\lambda}\tau^{a}_{\beta\eta}\hcd_{k_2+q_y,\beta}\hc_{k_2,\eta}:\nn
\cV_a(\bq,k_1,k_2)&=&\frac{v_a(\bq)}{L_xL_y}e^{-\frac{\ell^2q^2}{2}}e^{-i\ell^2q_x(k_1-k_2-q_y)}.
\eean

\subsection{Hartree Fock Approximation }

Now one assumes that the ground state is a single Slater determinant (SSD), which is translation invariant up to intervalley coherence. Any SSD is uniquely specified by its one-body averages 
\bean
\lgl\Psi|\hcd_{k,\beta}\hc_{k^\prime,\eta}|\Psi\rgl=\delta_{k,k^\prime}P_{\eta\beta},
\eean
The matrix $P$ is simply the projector to the occupied subspace in the four-dimensional spin/valley space. 

One now decomposes the two-body terms in the Hamiltonian to one-body terms by taking all possible averages, with the resulting mean-field Hamiltonian, including the Zeeman coupling $E_Z$ and the sublattice symmetry breaking (valley Zeeman) coupling $E_V$, being
\cite{Das_Kaul_Murthy_2022}
\bean\label{Hmf}
&&\hat H_{MF}[P]=\sum_{k_1}\Big(\sum_{a}\bM^{a}[P]-E_Z\sigma^z-E_V\tau^z\Big)_{\alpha\lambda}\hcd_{k_1,\alpha}\hc_{k_1,\lambda},\nn
&& \bM^{a}[P]_{\alpha\lambda}={u_{a,H}}{\rm Tr}(\tau^aP)\tau^a_{\alpha\lambda}
-{u_{a,F}}\big(\tau^aP \tau^a\big)_{\alpha\lambda},\\
&&  u_{a,H}=\frac{v_a({\mathbf 0})}{2\pi\ell^2},\ u_{a,F}=\int \frac{d^2\bq}{(2\pi)^2}v_a(\bq)e^{-\frac{\ell^2q^2}{2}},\label{uhuf}
\eean 

Correspondingly, we also have the ground state energy, which is a functional of the projector $P$. 
\bean\label{Ehf}
E[P]&=&\frac{1}{2}\sum_a\bigg[u_{a,H}{\rm Tr}\big(\tau^aP\big)^2-u_{a,F}{\rm Tr}\big(\tau^aP\tau^aP\big)\bigg]\nn
&&-E_Z{\rm Tr}(\sigma^zP)-E_V{\rm Tr}(\tau^zP).
\eean 
The minimization of $E(P)$ also diagonalizes $\hat H_{MF}$. In the USR limit, $v_a(\bq)=g_a$ will be independent of $\bq$, therefore, $u_{a,H}=u_{a,F}=\frac{g_a}{2\pi\ell^2}$. In the USR limit the phase diagram was obtained by Kharitonov in the parameter space $(g_\perp,g_z)$.
However, we will keep the discussion general, allowing $u_{a,H}\neq u_{a,F}$ in what follows.

\subsection{Variational Spinor Ansatz, Order Parameters}
At $\nu=0$, the $n=0$ LL manifold is half-filled, and  two out of 4 spinors are occupied. In this case the following ansatz for these 4 undetermined spinors completely exhausts all the possible ground states. 
\bean\label{f_ansatz}
&&\lvert f_1\rangle =\cos\frac{\alpha_1}{2}\lvert\btau,\bs_a\rangle+\sin\frac{\alpha_1}{2}\lvert-\btau,-\bs_b\rangle,\nn
&&\lvert f_2\rangle =\cos\frac{\alpha_2}{2}\lvert\btau,-\bs_a\rangle+\sin\frac{\alpha_2}{2}\lvert-\btau,\bs_b\rangle,\nn
&&\lvert f_3\rangle =\sin\frac{\alpha_1}{2}\lvert\btau,\bs_a\rangle-\cos\frac{\alpha_1}{2}\lvert-\btau,-\bs_b\rangle,\nn
&&\lvert f_4\rangle =\sin\frac{\alpha_2}{2}\lvert\btau,-\bs_a\rangle-\cos\frac{\alpha_2}{2}\lvert-\btau,\bs_b\rangle\nn
&&\lvert\btau\rangle=\begin{pmatrix}
    \cos\frac{\theta_{\tau}}{2}\\
    e^{i\phi_{\tau}}\sin\frac{\theta_{\tau}}{2}
\end{pmatrix},\quad \quad 
  \lvert\bs\rangle=\begin{pmatrix}
    \cos\frac{\theta_{s}}{2}\\
    e^{i\phi_{s}}\sin\frac{\theta_{s}}{2}
\end{pmatrix},
\eean
Here $|f_1\rangle,~|f_2\rangle$ are the occupied spinors, and $|\btau,\bs\rangle {\equiv} |\btau\rangle{\otimes}|\bs\rangle$, with $\btau$ ($\bs$) being unit vectors on the valley (spin) Bloch spheres. The $U(1)_{valley}$ symmetry allows us to choose $\phi_{\tau}{=}0,\pi$, which means we are working with real valley spinors. Similarly the $U(1)_{spin}$ symmetry of spin rotations around the direction of total field allows us to choose $\phi_{s}=0,\pi$. All four four-dimensional spinors can be chosen to be real.\\
The projector for filling $\nu$ can be constructed as 
\bean
P=\sum_{i=1}^{\nu+2}|f_i\rgl\lgl f_i|.
\eean 

To differentiate phases from each other, various order parameters $\hat O$'s are computed as 
\bean\label{norm_OP}
\langle\hat O\rangle=\frac{{\rm Tr}(P\hat O)}{2}.
\eean 
Since our spinors are all real, only real one-body operators can have a nonzero expectation value in the ground state. The order parameters $\hat O$ we study are expectation values of the following one-body operators, written schematically as $\sigma_z$ [representing ferromagnetic (FM) order], $\tau_z\sigma_x$ [representing canted antiferromagnetic (CAFM) order], $\tau_z$ [representing charge-density-wave (CDW) order in the ZLLs due to spin-valley locking], $\tau_x$ [representing intervalley coherence or bond order (BO), sometimes also termed Kekul\'e distorted (KD) order], $\tau_x\sigma_x$ [representing spin-valley entangled-X (SVEX) order], $\tau_y\sigma_y$ (SVEY), and $\tau_z\sigma_z$ [representing antiferromagnetic (AFM) order].

\section{Time Dependent Hartree Fock Formalism for Quantum Hall Ferromagnets}
\label{sec3}
Once the optimal ground state $|\Psi\rgl$ of mean-field Hamiltonian Eq.\eqref{Hmf} is obtained, we construct a unitary transformation to the basis that diagonalizes $H_{MF}$
\bean
&&\hc_{k,\alpha}=U_{\alpha\beta}\hd_{k,\beta},\ \hcd_{k,\alpha}=\hdd_{k,\beta}U^\dagger_{\beta\alpha},\\
&&\tilde \tau^a=U^\dagger \tau^a U,\  \tilde\sigma^a=U^\dagger \sigma^a U.
\eean
The spinors $|f_i\rgl=\hdd_i|0\rgl$ are precisely the eigenvectors of the $4\times4$ mean field Hamiltonian matrix, 
\bean\label{eigenvecotrs}
H_{MF}[P]|f_i\rgl=\epsilon_i|f_i\rgl
\eean 
or even more explicitly,
\bean
H_{MF}[P]=\underbrace{\begin{pmatrix}
   |f_1\rgl &|f_2\rgl&|f_3\rgl&|f_4\rgl
\end{pmatrix}}_{U}
\underbrace{\begin{pmatrix}
    \epsilon_1& & & \\
     & \epsilon_2& & \\
      & &\epsilon_3 & \\
       & & & \epsilon_4\\
\end{pmatrix}}_{\cE}
\underbrace{\begin{pmatrix}
   \lgl f_1| \\ \lgl f_2| \\ \lgl f_3| \\ \lgl f_4|
\end{pmatrix}}_{U^\dagger}.\nonumber
\eean 
Clearly, the projector $P$ is diagonal in the new basis, i.e.
\bean
P_{\lambda\alpha}=\lgl\Psi|\hdd_{\alpha}\hd_\lambda|\Psi\rgl=n_F(\alpha)\delta_{\alpha\lambda}.
\eean 
The eigenvalues of $H_{MF}$ using the rotated basis are  
\bean
\epsilon_\alpha&=&\sum_{a,\gamma}n_F(\gamma)\Big(u_{a,H} \tilde \tau^a_{\gamma\gamma}\tilde \tau^a_{\alpha\alpha}-u_{a,F}\tilde \tau^a_{\gamma\alpha}\tilde \tau^a_{\alpha\gamma}\Big)\nn
&&-E_Z\tilde\sigma^z_{\alpha\alpha}-E_V\tilde\tau^z_{\alpha\alpha}.
\eean 
We will always order the eigenvalues as $\epsilon_1\leq \epsilon_2<\epsilon_3\leq\epsilon_4$. By this choice, the transport gap $\Delta$ is given by 
\bean
\Delta=\epsilon_3-\epsilon_2.
\label{eq:Delta_definition}\eean

Now we are ready to consider the spectrum of collective excitations. Consider an excitation operator \cite{MSF2014,MSF2016,Murthy_Shankar_LLmix} in the new basis
\bean
\hat O_{\theta\phi}(\bq)=\sum_k e^{-iq_x(k-\frac{q_y}{2})\ell^2}\hdd_{k-q_y,\theta}\hd_{k,\phi}.
\eean 
The time evolution of this excitation follows the Heisenberg equation of motion,
\bean\label{Heisenberg_EOM}
-i\frac{d\hat O_{\theta\phi}(\bq)}{dt}=\big[\hat H,\hat O_{\theta\phi}(\bq)\big].
\eean 
The commutator above consists of both one-body and two-body terms. In the TDHF approximation,  the two-body terms are reduced to one-body terms by HF expectation values. This leads to the following closed set of equations in the 4-dimensional 1-exciton subspace: 
\bean
&&\big[\hat H,\hat O_{\theta\phi}(\bq)\big]_{HF}=\Big(\epsilon_\theta-\epsilon_\phi\Big)\hat O_{\theta\phi}(\bq)\nn
&&\quad+\Big (n_F(\phi)-n_F(\theta)\Big)\times\nn
&&\quad\quad\bigg[\bv_{a,H}(\bq)\tilde \tau^a_{\phi\theta}\tilde \tau^a_{\alpha\lambda}-\bv_{a,F}(\bq)\tilde\tau^a_{\alpha\theta}\tilde \tau^a_{\phi\lambda}\bigg]\hat O_{\alpha\lambda}(\bq),\label{TDHF}\\
&&\bv_{a,H}(\bq)=\frac{v_a(\bq)}{2\pi\ell^2},\nn
&&\bv_{a,F}(\bq)=\int \frac{d^2\bp}{(2\pi)^2}e^{-\frac{p^2\ell^2}{2}}v_a(\bp)e^{i\hat z\cdot(\bq\times\bp)\ell^2}.\label{vqhf}
\eean 
The TDHF collective modes $\omega(\bq)$ are obtained by diagonalizing this matrix at each $\bq$.  \\

\section{Phase Diagrams with non-USR interaction}
\label{sec4}

A powerful way to characterize two-body interactions within a particular Landau level is by means of Haldane pseudopotentials $V_m$, which represent the amplitude of the interaction in the relative angular momentum $m$ channel of the two electrons. Within the $n=0$ LL manifold, a USR interaction corresponds to all  $V_m=0$ except for $m=0$. To go beyond this assumption in a minimal way, while keeping the interactions short-ranged, we now allow $V_0\neq 0,\ V_1\neq0$, while $V_m=0, m>1$. This leads to the  $v_i(\bq)$ of Eq. \eqref{eq:vaq_def} taking the following form
\bean\label{viq}
v_a(\bq)=g_a\Big(1+r_aL_1(q^2\ell^2)\Big)
\eean 
where $L_1(q^2\ell^2)$ is the Laguerre polynomial and $r_a$ is a dimensionless number describing the ratio of interaction amplitudes $V_{a,1}/V_{a,0}$. Microscopically, these slightly longer-range interactions arise from integrating out the higher LL-manifolds via LL-mixing~\cite{Das_Kaul_Murthy_2020}. One might wonder whether adding nonzero pseudopotentials $V_m,~m>1$ would result in more phases. It turns out that for $\nu=0$ having nonzero $V_0,~V_1$ is sufficient to exhibit all the potential phases. 

\begin{widetext}  

\begin{figure}[H]
    \centering
  \includegraphics[width=0.95\textwidth,height=0.59\textwidth]{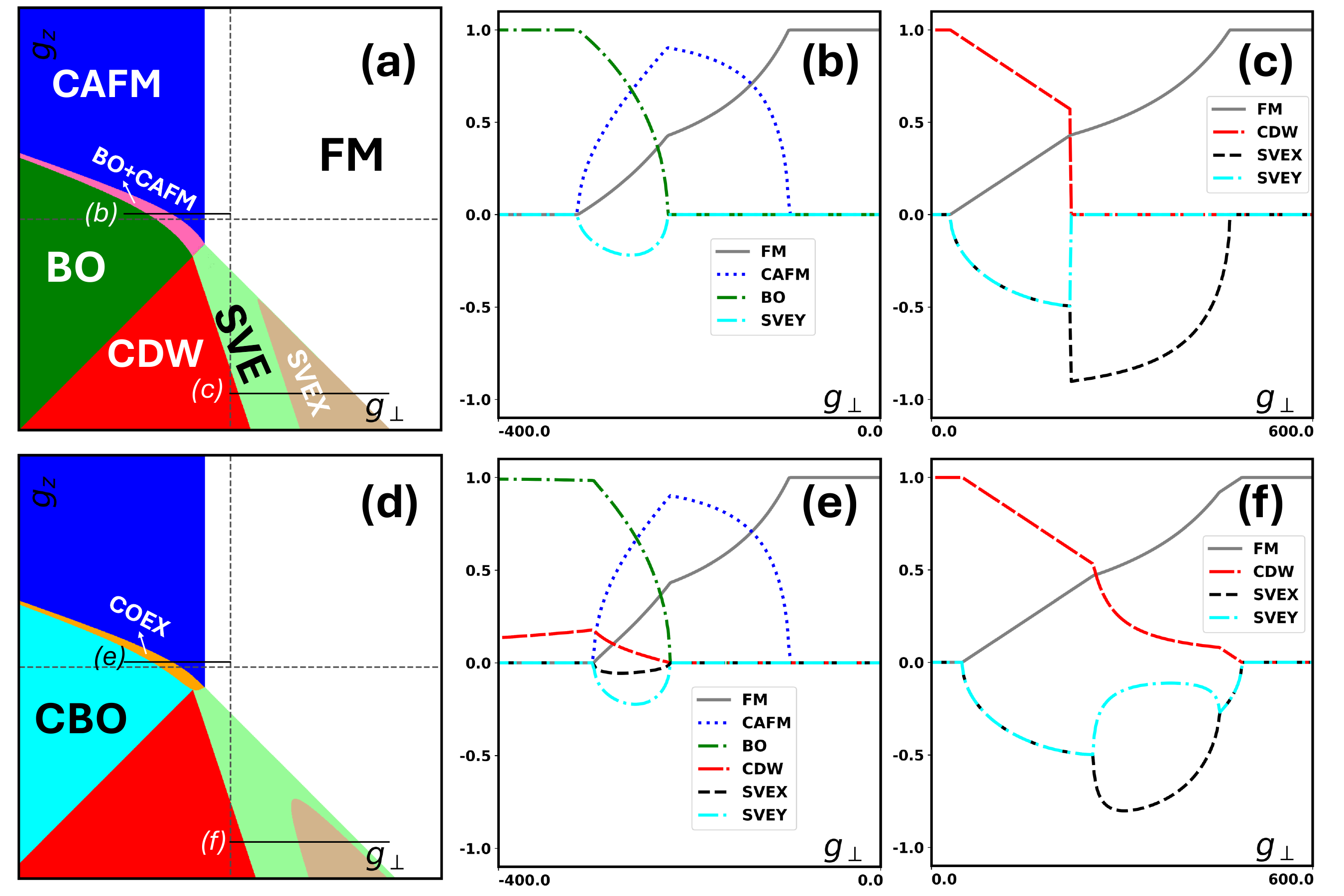}
    \caption{$\nu=0$ phase diagrams with $ g_{\perp}\in [-800,800]\text{meV}\cdot \text{nm}^{2}$, $ g_{z}\in [-800,800]\text{meV}\cdot \text{nm}^{2}$ and the ratios are fixed at $r_\perp=r_z=-0.2$ and (a) $E_Z=1e_Z,\ E_V=0$ (d) $E_Z=1e_Z,\ E_V=0.1e_Z$. (b-c) Relevant order parameters along horizontal sections in (a). (e-f) Relevant order parameters along horiontal sections in (d). }
    \label{nn_diagram}
\end{figure}
\end{widetext}

Note that there are four independent couplings, making the full space of parameters very hard to visualize. In what follows, we keep $r_z,~r_\perp$ fixed while varying $g_a$. This corresponds to taking a two-dimensional section through the four-dimensional coupling constant space. A microscopic estimate for the interacting strength  has recently been carried out~\cite{Wei_Xu_Sodemann_Huang_LLM_SU4_breaking_MLG_2024}, with the results being
\bean
g_\perp=-26 \text{ meV$\cdot$ nm}^{2},\ g_z=184 \text{ meV$\cdot$ nm}^{2}.
\eean 
We will present phase diagrams in the full range of couplings, including regions very far from the estimate presented above. From \eqref{uhuf}, the Hartree and Fock couplings are given by  
\bean
&&u_{\perp,H}=\frac{g_\perp}{2\pi\ell^2}\Big(1+r_\perp\Big),\quad u_{\perp,F}=\frac{g_\perp}{2\pi\ell^2}\Big(1-r_\perp\Big),\nn  
&&u_{z,H}=\frac{g_z}{2\pi\ell^2}\Big(1+r_z\Big),\quad u_{z,F}=\frac{g_z}{2\pi\ell^2}\Big(1-r_z\Big).
\eean 
Keeping the ratios $r_i$ constant keeps the ratios of the Hartree and Fock couplings constant.  In Fig. \ref{nn_diagram}, we present the phase diagram for   $r_\perp=r_z=-0.2$  in $(g_\perp,\ g_z)$ space for the small $E_V$. Throughout this paper, we choose to fix the magnetic field at $B_{\perp}{=}10~{\rm Tesla}$, implying a Zeeman energy of $e_Z{=}(1/2)g_{B}\mu_B B_{\perp}{=}0.58~{\rm meV}$ [the $g$-factor $g_{B}{=}2$ for graphene]. In later phase diagrams, we will take this $e_Z$ as the energy unit, i.e. $e_Z{=}1$. 
Out of the phases presented in Fig. \ref{nn_diagram}(a) and (d), the ferromagnetic (FM), (canted) antiferromagnetic ((C)AFM), (canted) bond order ((C)BO), charge density wave (CDW) phases can be found in the USR limit\cite{kharitonov2012:nu0}. The three other phases- BO+CAFM, SVE, and SVEX - only appear when $r_a\neq0$.  

The phases are distinguished from each other by the symmetries they break,  and thus  the nonzero order parameters they manifest. In Figs.~\ref{nn_diagram}(b) and (c), we present the order parameters along two different cuts in Fig.~\ref{nn_diagram}(a), shown as black lines. Recall that for the top row of panels $E_V=0$. The cut in Fig.~\ref{nn_diagram}(b) for small positive $g_z$ shows the system starting in the BO phase, from which is makes a II-order transition into the BO+CAFM phase, followed by another II-order transition into the CAFM phase. One can see that the BO+CAFM phase has both magnetic and lattice symmetry breaking. Eventually, for large enough $g_\perp$ the system goes into the FM phase. The second cut at  large negative $g_z$ starts in the CDM phase, makes a II-order transition into the SVE phase, and then a I-order transition into the SVEX phase, from which it eventually goes into the FM phase. The lower set of panels shows what happens when a tiny $E_V=0.1$ is turned on. The entire BO phase acquires a nonzero CDW order parameter, turning it into the canted BO (CBO) phase. Similarly, the BO+CAFM phase turns into the COEX phase. Last, but not least, the I-order transition between the SVE and SVEX phases changes into a II-order transition, and the SVEX phase acquires a small SVEY order parameter as well. We will continue to call it SVEX in what follows to keep the notation simple. 

More detailed information about the state is provided by the eigenvectors of the mean-field Hamiltonian as functions of $u_{a,H},\ u_{a,F},\ E_Z,\ E_V$.

It is worth noting that the $\nu=0$ quantum Hall ferromagnetic state is a singlet under transformations among the occupied states, in the sense that we can take linear combinations between $|f_1\rgl$ and $|f_2\rgl$ with arbitrary angles $\eta_1$ and $\eta_2$
 \bean\label{linear_combination}
&&\begin{pmatrix}
   |f_1\rgl \\
    |f_2\rgl\\
\end{pmatrix}\ \rightarrow\ 
\begin{pmatrix}
   \cos\frac{\eta_1}{2} &\sin\frac{\eta_1}{2}\\
    -\sin\frac{\eta_1}{2}& \cos\frac{\eta_1}{2}\\
\end{pmatrix}
\begin{pmatrix}
   |f_1\rgl \\
    |f_2\rgl\\
\end{pmatrix},\nn
&&\begin{pmatrix}
   |f_3\rgl \\
    |f_4\rgl\\
\end{pmatrix}\ \rightarrow\ 
\begin{pmatrix}
   \cos\frac{\eta_2}{2} &\sin\frac{\eta_2}{2}\\
    -\sin\frac{\eta_2}{2}& \cos\frac{\eta_2}{2}\\
\end{pmatrix}
\begin{pmatrix}
   |f_3\rgl \\
    |f_4\rgl\\
\end{pmatrix}.
\eean 
 This basis change keeps the projector $P$ to the occupied states invariant. However, the basis which diagonalizes $H_{MF}$ picks out two particular eigendirections in the two-dimensional space, which are the spinors we will present shortly. 
 
 Below we list the eigenvectors of $H_{MF}$ in each phase, along with the ground state energies, and their identifying color codes which we will use throughout this work.
\bean
\tikz{ \draw[draw=black, fill=white] (0,0) circle (0.75ex)}\ {\rm FM:}\ &&|f_1\rgl=|K,\uparrow\rgl,\ |f_2\rgl=|K^\prime,\uparrow\rgl,\nn
&&E^{{\rm FM}}=-2\big(E_Z+u_{\perp,F}\big)-u_{z,F}.
\eean 
\bean
\tikz{ \draw[draw=black, fill=blue] (0,0) circle (0.75ex)}\ {\rm CAFM:}\ &&|f_1\rgl=|K,\nwarrow_s\rgl,\ |f_2\rgl=|K^\prime,\nearrow_s\rgl,\nn
&&|\nwarrow_s\rangle=\begin{pmatrix}
    \cos\frac{\theta_s}{2}\\
    \sin\frac{\theta_s}{2}
\end{pmatrix},\quad
|\nearrow_s\rangle=\begin{pmatrix}
    \cos\frac{\theta_s}{2}\\
    -\sin\frac{\theta_s}{2}
\end{pmatrix},\nn
&&\cos\theta_s=-\frac{E_Z}{2u_{\perp,F}},\nn
&&E^{{\rm CAFM}}=\frac{E_Z^2}{2u_{\perp,F}}-u_{z,F}.
\label{eq:angle_CAFM}\eean 
When $E_Z=0$ (a very idealized and unphysical limit), the CAFM phase simplifies to AFM as seen in Fig. \ref{Ev}(a). 
\bean
\tikz{ \draw[draw=black, fill=red] (0,0) circle (0.75ex)}\ {\rm CDW:}\ &&|f_1\rgl=|K,\uparrow\rgl,\ |f_2\rgl=|K,\downarrow\rgl,\nn
&&E^{{\rm CDW}}=-2E_V-u_{z,F}+2u_{z,H}.
\eean 
\bean
\tikz{ \draw[draw=black, fill=CBOcolor] (0,0) circle (0.75ex)}\ {\rm CBO:}\ &&|f_1\rgl=|\nwarrow_{\tau},\uparrow\rgl,\ |f_2\rgl=|\nwarrow_{\tau},\downarrow\rgl,\nn
&&|\nwarrow_\tau\rangle=\begin{pmatrix}
    \cos\frac{\theta_\tau}{2}\\
    \sin\frac{\theta_\tau}{2}
\end{pmatrix},\nn 
&&\cos\theta_\tau=\frac{E_V}{2u_{z,H}-u_{z,F}-2u_{\perp,H}+u_{\perp,F}},\nn
&&E^{\rm CBO}=2u_{\perp,H}-u_{\perp,F}-E_V\cos\theta_\tau.
\label{eq:angle_CBO}\eean 
This state interpolates smoothly between the BO and CDW phases. Clearly, at $E_V=0$, $|\nwarrow_{\tau}\rgl=\big(\frac{1}{\sqrt 2},\ \frac{1}{\sqrt 2}\big)^T\equiv |\hat e_x\rgl$, which is the BO state, while at large $E_V$, $|\nwarrow_{\tau}\rgl\to|K\rgl$, which is the CDW state. This implies  that the first-order phase transition from CDW to BO at $E_V=0$  becomes a second-order transition from CDW toCBO for any nonzero $E_V$. This is quite analogous to the first-order transition between the AFM and the FM phases at $E_Z=0$ becoming a second-order transition between the CAFM and FM phases for $E_Z>0$.\\
When the USR constraint is relaxed, the imbalance between Hartree and Fock couplings leads to phases which are spin valley entangled (SVE). In other words, the eigenvectors of $H_{MF}$ in these phases cannot be expressed as a direct product of a spin Bloch vector and a valley Bloch vector. This leads to the order parameters $\lgl\tau_x\sigma_x\rgl$ or $\lgl\tau_y\sigma_y\rgl$ becoming nonzero in these phases. The appearance of these SVE phases depends on the ratios $\frac{u_{a,F}}{u_{a,H}}$.  As discussed in \cite{Das_Kaul_Murthy_2022,De_etal_Murthy2022,Stefanidis_Sodemann2023}, the maximum number of phases are observed when  $\frac{|u_{\perp,F}|}{|u_{\perp,H}|}>1\ \&\  \frac{|u_{z,F}|}{|u_{z,H}|}>1$. This can be realized by choosing $r_\perp=r_z=-0.2$, which is what we chose in Fig. \ref{nn_diagram}. The observed SVE phases can be categorized into three types.

\begin{figure}[H]
    \centering
  \includegraphics[width=0.485\textwidth,height=0.48\textwidth]{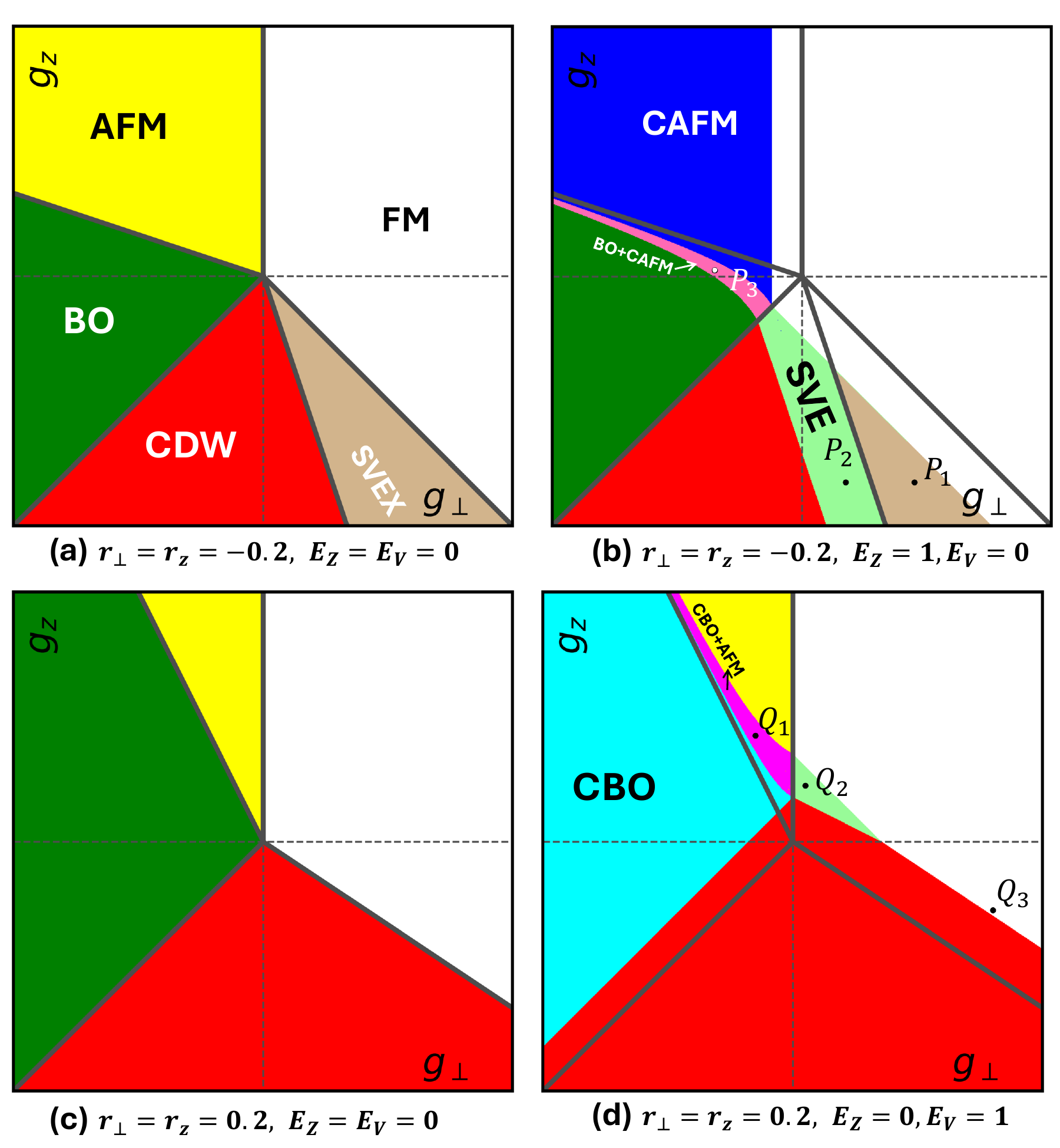}
    \caption{Phase diagrams with and without spin and valley Zeeman fields. The ranges of the couplings are $ g_{\perp}\in [-800,800]\text{meV}\cdot \text{nm}^{2}$, $ g_{z}\in [-800,800]\text{meV}\cdot \text{nm}^{2}$ with fixed ratios $r_\perp,\ r_z$. (a) $E_Z=E_V=0$ and $r_z,r_\perp<0$. All phases are degenerate at the origin, and depend only on the ratio $g_\perp/g_z$. Five phases are present. The longer range interactions make the SVEX phase possible. (b) The same ratios as in panel (a), but not the Zeeman coupling is nonzero whereas the valley Zeeman is still zero. The AFM phase changes to the CAFM phase, and two new phases appear due to $E_Z>0$, the BO+CAFM phase and the SVE phase. To highlight the differences, we also plot the solid black lines which indicate where the phase boundaries were in panel (a) at $E_Z=0$. Three points $P_{1,2,3}$ are highlighted for exploring the variation of order parameters and 1-body energies with increasing $E_Z$.
    (c) The phase diagram at $E_Z=E_V=0$ and $r_z,r_\perp>0$. Now only four phases appear, the same as in the USR case. (d) Upon adding $E_V>0$ while keeping $E_Z=0$ several changes occur. The BO phase acquires a valley polarization and changes to the CBO phase. The CBO+AFM phase appears near the old BO-AFM boundary, and the SVE phase appears near the old CDW-FM boundary. Once again, the solid black lines indicate the phase boundaries of panel (c). Three points $Q_{1,2,3}$ have been marked for further exploration with increasing $E_V$.}
    \label{Ez_Ev}
\end{figure}

\begin{figure}[H]
    \centering
  \includegraphics[width=0.48\textwidth,height=0.56\textwidth]{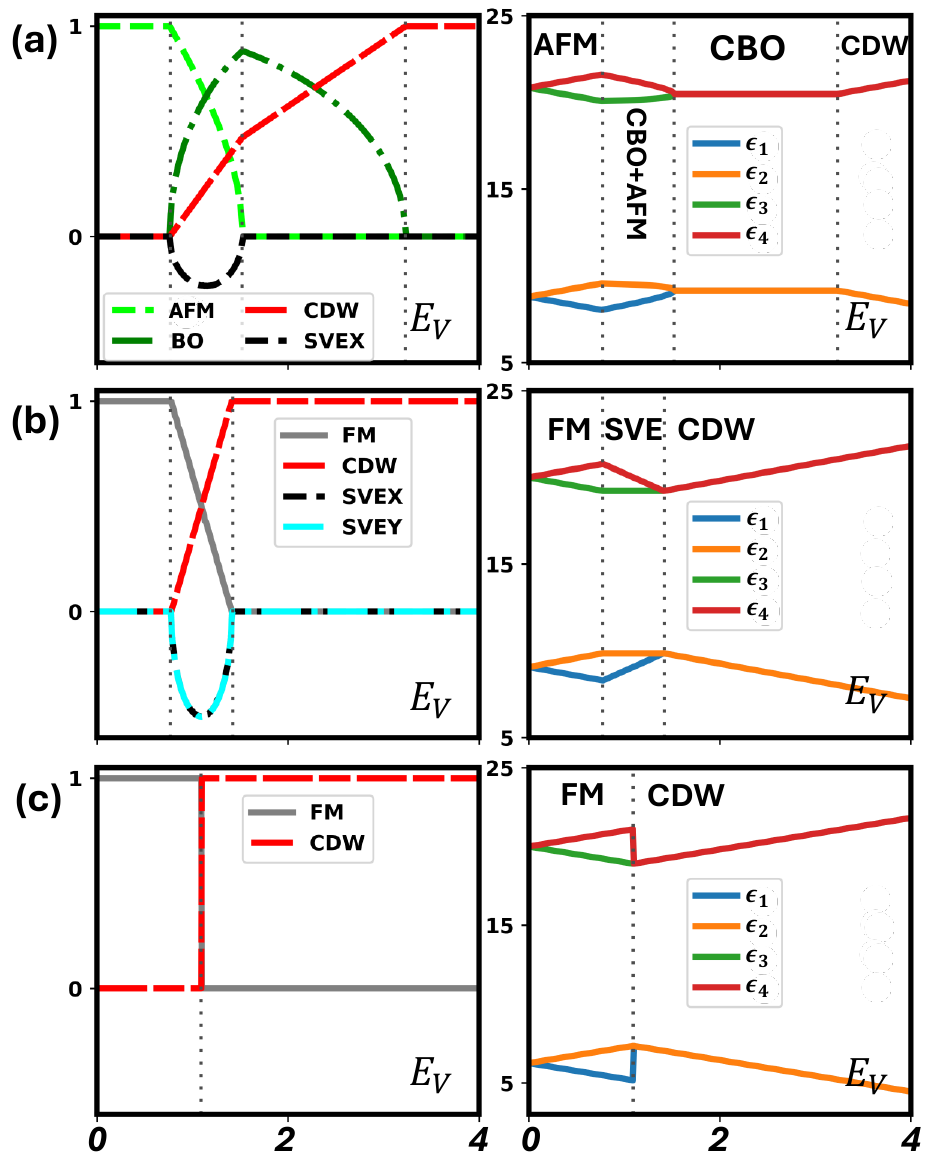}
    \caption{Development order parameters (in left panels) and one-body eigenvalues (in right panels) with $E_V$ increasing from 0 to 4 for points  (a) $Q_1$ (b)$Q_2$  (c)$Q_3$  highlighted in Fig. \ref{Ez_Ev}(d).}
    \label{Ev}
\end{figure}

\begin{itemize}[left=0pt]
    \item \tikz{ \draw[draw=black, fill=SVEcolor] (0,0) circle (0.75ex)} Spin-Valley-Entangled Phase (SVE):\\
\bean\label{SVE_spinors}
&&|f_1\rgl=\sin\frac{\alpha}{2}|K,\downarrow\rgl-\cos\frac{\alpha}{2}|K^\prime,\uparrow\rgl,\quad |f_2\rgl=|K,\uparrow\rgl,\nn
&&\cos\alpha=\frac{E_Z-E_V+u_{z,H}+u_{\perp,F}}{u_{z,H}-u_{z,F}}.
\label{eq:angle_SVE}\eean 
This phase appear if one turns on a nonzero $E_Z$ or $E_V$. A key observation is that since there is only 1 spin-valley entangled spinor, there is a relation between the SVEX and SVEY order pamameters: $\lgl\tau_x\sigma_x\rgl=\lgl\tau_y\sigma_y\rgl$.
    \item \tikz{ \draw[draw=black, fill=SVExcolor] (0,0) circle (0.75ex)} Spin-Valley-Entangled-x Phase (SVEX):\\
\bean\label{E-BO-AFM_spinors}
&&|f_1\rgl=\cos\frac{\alpha_1}{2}|K,\downarrow\rgl-\sin\frac{\alpha_1}{2}|K^\prime,\uparrow\rgl,\nn 
&&|f_2\rgl=\cos\frac{\alpha_2}{2}|K,\uparrow\rgl-\sin\frac{\alpha_2}{2}|K^\prime,\downarrow\rgl\nn
&&\cos\alpha_{1/2}=\frac{E_V}{2u_{z,H}-u_{z,F}+u_{\perp,F}}\pm\frac{E_Z}{u_{\perp,F}+u_{z,F}}.
\label{eq:angle_SVEX}\eean 
It is instructive to take various limiting cases. Specifically, at $E_Z=E_V=0$ (see Fig. \ref{z0v0}), the spinors simplify to 
\bean\label{svex_z0v0}
\frac{1}{\sqrt 2}\Big(|K,\dwa\rgl-|K^\prime,\uparrow\rgl\Big),\quad \frac{1}{\sqrt 2}\Big(|K,\uparrow\rgl-|K^\prime,\dwa\rgl\Big)
\eean 
which has SVEX order paramter $\lgl\tau_x\sigma_x\rgl=-1$ being nonzero but SVEY vanishing. 
For $E_Z>0,~E_V=0$,  one obtains $\cos\alpha_1=-\cos\alpha_2$.  The phase now has a  nonzero FM order parameter $\lgl\sigma_z\rgl$. Similarly, for $E_Z=0,~E_V>0$, one obtains $\cos\alpha_1=\cos\alpha_2$. This implies that the phase has  a nonzero CDW order parameter $\lgl\tau_z\rgl$.
    \item \tikz{ \draw[draw=black, fill=coexistcolorzv] (0,0) circle (0.75ex)} Coexistence (COEX) of (canted) Bond Order and (canted) Antiferromagnetic ((C)BO+(C)AFM):\\
    \bean\label{CBO_CAF_spinors}
&&|f_1\rgl=\cos\frac{\alpha_1}{2}|\btau,\uparrow\rgl-\sin\frac{\alpha_1}{2}|-\btau,\downarrow\rgl,\nn
&&|f_2\rgl=\cos\frac{\alpha_2}{2}|\btau,\downarrow\rgl-\sin\frac{\alpha_2}{2}|-\btau,\uparrow\rgl.
\eean 
In the spinors above, there are 3 nontrivial angles $\alpha_1$, $\alpha_2$ and the canting  angle $\theta_\tau$ on the valley Bloch sphere. Unfortunately, these angles  can only be obtained  numerically in the general case, so we are not able to present analytical expressions. However, in extremal cases, the angles can be found analytically. For $E_V=0$ we have $\btau=\hat e_x$ which is just the \BCcircle\ BO+CAFM phase in Fig.\ref{nn_diagram}(a) (Analytically expressions of $\alpha_1\ \&\ \alpha_2$ are given in Appendix \ref{app: C}). For $E_Z=0$ we have $\alpha_1=\alpha_2$,  which is just the \CAcircle\ CBO+AFM in Fig.\ref{Ez_Ev}(d). 
\end{itemize}

 In Fig.~\ref{Ez_Ev}(a), we present the four different phase diagrams designed to show the phases that appear when the ratios $r_z,r_\perp$ change, and also the interesting fact that certain phases only appear in the presence of the 1-body couplings $E_Z,~E_V$. We only show the cases when both $r_z,r_\perp$ positive or both negative, because these cases exhibit all the possible phases. Phase diagrams for other signs of the ratios are presented in Appendix \ref{app: A}. Fig.~\ref{Ez_Ev}(a) shows the phase diagram   for $E_V=E_Z=0$ and $r_z,~r_\perp<0$. As expected, all the phases converge at the noninteracting point $g_z=g_\perp=0$. As compared to Kharitonov's USR phase diagram, only the SVEX phase is new. However, as shown by Fig.~\ref{Ez_Ev}(b), the situation changes when $E_Z>0$ while keeping $E_V=0$. The CAFM phase acquires a nonzero spin polarization and becomes the CAFM phase. Two additional phases appear, the BO+CAFM phase near the boundary of the AFM and BO phases at $E_Z=0$, and the SVE phase between the SVEX and CDW phases. This highlights the role of externally tunable 1-body potentials in facilitating the presence of new phases, and changing the topology of the phase diagram. Fig.~\ref{Ez_Ev}(c) presents the phase diagram for $E_Z=E_V=0$ but now with $r_z,~r_\perp>0$. The phase diagram is very similar to the USR case. However, upon turning on $E_V>0$, significant changes occur in the phase diagram, as shown in Fig.~\ref{Ez_Ev}(d). The BO phase acquires a valley polarization and becomes the CBO phase. Additionally, two new phases appear, the CBO+AFM phase near the old BO-AFM phase boundary, and the SVE phase near the CDW-FM phase boundary. Since one of the main goals of this work is to explore the evolution of observables as the 1-body couplings $E_Z,~E_V$ are varied, three points have been chosen in each of panels (b) and (d) for further exploration, to which we now turn.

\begin{figure}[H]
    \centering
  \includegraphics[width=0.48\textwidth,height=0.56\textwidth]{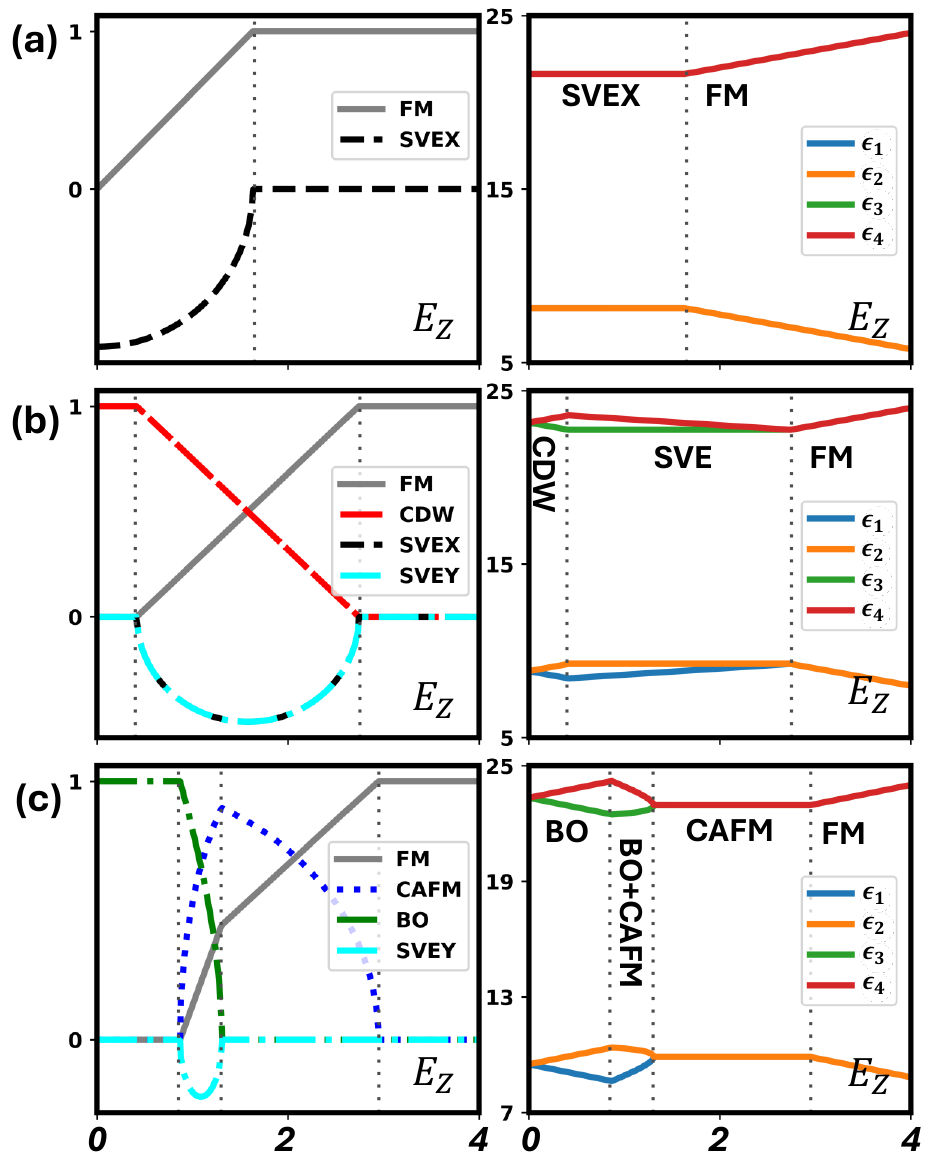}
    \caption{Development order parameters (in left panels) and one-body eigenvalues (in right panels) with $E_Z$ increasing from 0 to 4 for points  (a) $P_1$ (b)$P_2$  (c)$P_3$  highlighted in Fig. \ref{Ez_Ev}(b).}
    \label{Ez}
\end{figure}

 Let us first consider the evolution of the physical properties of the phases as $E_Z$ increases. In Figs.~\ref{Ez}(a)-(c) we show the order parameters versus $E_Z$ starting from the points $P_1,~P_2,~P_3$ respectively chosen in Fig.~\ref{Ez_Ev}(b). We assume that $E_V=0$ to focus on the behavior with $E_Z$ alone.   Qualitatively, as $E_Z$ increases, all the phases slide diagonally downwards, as the FM phase, which is energetically favored the most by $E_Z$, takes up more room in the coupling constant space. The behavior of  the order parameters is presented in the left panels, while the behaviour of  the eigenvalues of $H_{MF}$ are presented in the corresponding right panels. In some cases the eigenvalues are degenerate. The most physically relevant panel for graphene is Fig.~\ref{Ez}(c). The system starts in the BO phase at $E_Z=0$ and evolves through the COEX and CAFM phases before ending in the FM phase. We note that the point $P_3$ is in the COEX phase in the phase diagram of Fig.~\ref{Ez_Ev}(b), but this is because the phase diagram is at $E_Z=1$ rather than $E_Z=0$. From the right panel of Fig.~\ref{Ez}(c) one can also infer the behavior of the transport gap, which is the difference between the energies of the highest occupied and the lowest unoccupied state. The transport gap decreases in the BO phase, increases in the BO+CAFM phase, and remains constant with $E_Z$ in the CAFM phase. Though not shown, for very large $E_Z$ the system goes over into the FM phase, where the transport gap once again increases with $E_Z$. It can be seen that this indeed corresponds to the evolution shown earlier in Fig.~\ref{coex_gap}. \\
 
 Next we turn to the behavior of the order parameters and 1-body eigenvalues of $H_{MF}$ versus $E_V$, while keeping $E_Z=0$. This is an unrealistic limit, but one can approach this limit in real samples by making $E_V,|g_a|>>E_Z$, because typically $E_Z$ is the smallest energy scale in the problem. Qualitatively, seen in Fig.~\ref{Ez_Ev}(d), as $E_V$ increases, the entire phase diagram shifts upwards, because the CDW phase, which is energetically the most favored by $E_V$, will take over more room in the coupling constant space. In Fig.~\ref{Ev}, rows (a), (b), and (c) of panels correspond to the points $Q_1,~Q_2,~Q_3$ respectively, which were chosen in Fig.~\ref{Ez_Ev}(d). The left panels of Fig.~\ref{Ev}(a)-(c) show plots of the order parameters versus $E_V$, while the right panels show the corresponding HF eigenvalues. The most physically relevant plot for graphene is Fig.~\ref{Ev}(a). The system starts in the AFM phase at $E_V=0$, makes a second-order transition into the CBO+AFM phase as $E_V$ increases, and then another second-order transition into the CBO phase, and finally ends up in the CDW phase for large $E_V$. The behavior of the 1-body eigenvalues of $H_{MF}$ is shown in the right panel of Fig.~\ref{Ev}(a). It can be seen that the gap first decreases in the AFM phase, increases in the CBO+AFM phase, remains constant in the CBO phase, and eventually increases again with $E_V$ when the system reaches the CDW phase. 

 Now we are ready to examine the collective modes and their dispersions in the various phases. 

\section{Transport Gaps and Collective Excitations}
\label{sec:Gaps_Collective_Excitations}
In this section we will first systematically present the behavior of the transport gaps in each phase as a function of the 1-body couplings $E_Z,~E_V$. We will then go on to describe the collective excitation spectrum in each phase. 

First we need a few preliminaries. 
Given the HF projector matrix for the phases observed in the space of coupling constants, we can perform the TDHF calculation to explore their collective excitations. From Eqs.\eqref{Heisenberg_EOM} and \eqref{TDHF}, we know the eigenoperators satify
\bean
&&\omega(\bq)\hat O_{\theta\phi}(\bq)=\sum_{\alpha\lambda}\cK_{\theta\phi,\alpha\lambda}(\bq)\hat O_{\alpha\lambda}(\bq),\nn
&&\cK_{\theta\phi,\alpha\lambda}(\bp)=\Big(\epsilon_\theta-\epsilon_\phi\Big)\delta_{\theta\alpha}\delta_{\phi\lambda}\nn
&&\quad\quad\quad\quad\quad+\Big (n_F(\phi)-n_F(\theta)\Big)\times\nn
&&\quad\quad\quad\quad\bigg[\bv_{a,H}(\bq)\tilde \tau^a_{\phi\theta}\tilde \tau^a_{\alpha\lambda}-\bv_{a,F}(\bq)\tilde\tau^a_{\alpha\theta}\tilde \tau^a_{\phi\lambda}\bigg].
\eean 
Recall that we chose to relax the USR condition in a minimal way by assuming that the anisotropic short-range couplings have only the Haldane pseudopotentials $V_0,~V_1$ being nonzero. Therefore, from  \eqref{uhuf} and \eqref{viq}, we obtain
\bean
\bv_{a,H/F}(\bq)=e^{-\frac{q^2\ell^2}{2}}
\frac{g_a}{2\pi\ell^2}\Big(1\pm r_aL_1(q^2\ell^2)\Big).
\eean 

Besides the anisotropic matrices labeled by $a=\perp, z$, we will also include a USR density-density interaction to provide a spin stiffness\cite{MSF2014}, i.e.
\bean
\bv_0(\bq)=\bv_{0,H}(\bq)=\bv_{0,F}(\bq)=e^{-\frac{q^2\ell^2}{2}}u_0,
\eean 
where $u_0=\frac{g_0}{2\pi\ell^2}$ and $g_0$ will be taken to be much larger than $g_\perp, g_z$. Strictly speaking, the correct isotropic interaction to use is the Coulomb interaction. However, our focus in this paper will be on the low-lying long-wavelength ($q\to0)$ collective mode dispersions, whose qualitative features (such as gaplessness) are independent of the precise form of the isotropic interaction. Throughout this whole work, we will take the short-range isotropic interaction to have the strength $u_0=10e_Z$.
As illustrated in Fig. \ref{graphene_LLs}, at $\nu=0$, we need to consider all possible particle-hole operators which are labeled by the following pairs: $\big(1\rightarrow 4\big),\ \big(2\rightarrow 3\big),\ \big(4\rightarrow 1\big),\ \big(3\rightarrow 2\big),\ \big(1\rightarrow 3\big),\ \big(2\rightarrow 4\big),\ \big(3\rightarrow 1\big),\ \big(4\rightarrow 2\big)$. In these basis, we found that for most of the phases presented above, the TDHF matrix $\cK$ will be block diagonalized, 
\bean\label{block_TDHF}
\cK(q)=\begin{pmatrix}
   \bK^u(q) &\\
    & \bK^l(q)\\
\end{pmatrix},
\eean 
where the basis of $\bK^u$ will be  $\big(1\rightarrow 4\big),\ \big(2\rightarrow 3\big),\ \big(4\rightarrow 1\big),\ \big(3\rightarrow 2\big)$ (labelled by red curved arrows in Fig.~\ref{graphene_LLs}) while $\big(1\rightarrow 3\big),\ \big(2\rightarrow 4\big),\ \big(3\rightarrow 1\big),\ \big(4\rightarrow 2\big)$ (labelled by blue curved arrows in Fig.~\ref{graphene_LLs}) for $\bK^l$. In the following sections, we will present the TDHF excitation spectra for all observed $\nu=0$ QHFM phases.

Throughout this section, we consider the physical case $E_Z>0$  which means that the symmetry of the Hamiltonian is $U(1)_s\otimes U(1)_v$. 

\subsection{Simple States}
In this subsection we present the response of the transport gaps to external fields $E_Z$ and $E_V$ in the "simple" phases of which the filled spinors $|f_1\rgl,\ |f_2\rgl$ are direct products of valley and spin Bloch vectors. These are the FM,CAFM, CBO, and CDW states. We will also present the explicit form of their eigenvectors and  eigenvalues of $H_{MF}[P]$. Finally, we will present the dispersions of the collective modes of the simple states.

\begin{figure}[h]
    \centering \includegraphics[width=0.48\textwidth,height=0.24\textwidth]{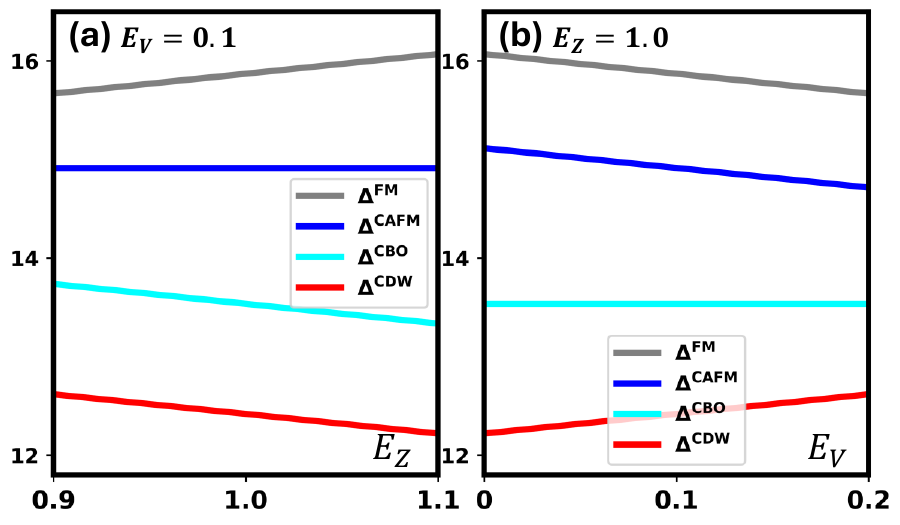}
    \caption{Response of transport gaps in phases FM, CAFM, CBO and CDW states to external 1-body couplings. (a) $E_Z$ 
 is increasing from 0.9 to 1.1 at fixed $E_V=0.1e_Z$. (b) $E_V$ 
 is increasing from 0 to 0.2 at fixed $E_Z=1e_Z$. The ratios are $r_z=r_\perp=-0.2$, as in  Fig.~\ref{nn_diagram}(d). The specific values of $g_z,~g_\perp$ used at $E_Z=0.9$ are different for the different phases.  The analytical results for the gaps in Eqs.~\eqref{fm_gap},~\eqref{cdw_gap},~\eqref{cbo_gap}, and \eqref{caf_gap} confirm that these results are generic, independent of precisely where one starts in a particular phase.  }
    \label{simple_gaps}
\end{figure}
\begin{itemize}[left=0pt]
    \item \FMcircle\ FM: Eigenvectors and eigenvalues of $H_{MF}[P^\text{FM}]$ are listed below, with the eigenvectors being the columns of the unitary matrix that transforms from the original basis to the basis diagonalizing $H_{MF}$.
\bean
&&U^\text{FM}=\ 
\begin{blockarray}{ ccccc }
|f_1\rgl &|f_2\rgl &|f_3\rgl &  |f_4\rgl & \\
& & & & &\\
\begin{block}{(cccc)c} 
  1 & 0 & 0 & 0 &  \\
  & & & & &\\
 0 & 0 & 1 & 0 &  \\
 & & & & &\\
 0 & 1 & 0 & 0 & \\
 & & & & & \\
 0 & 0 & 0 & 1 &   \\
\end{block}
\end{blockarray},\\
&&\epsilon_1=u_0-E_Z-E_V-2\unf-\uzf,\nn
&&\epsilon_2=u_0-E_Z+E_V-2\unf-\uzf,\nn
&&\epsilon_3=2u_0+E_Z-E_V,\nn
&&\epsilon_4=2u_0+E_Z+E_V.
\eean 
The gap between the highest filled and lowest unfilled one-body HF eigenvalues is the transport gap.
\bean\label{fm_gap}
\Delta^\text{FM}=u_0+2E_Z-2E_V+2\unf+\uzf.
\eean 
This shows that $\Delta^\text{FM}$ increases with $E_Z$ and decreases with $E_V$. The collective modes of the FM phase are shown in Fig.~\ref{fig:FM_excitations}. Two features are noteworthy: There are no gapless collective modes, because the phase does not break any continuous symmetries. Secondly, any phase with a nonzero spin polarization and $SU(2)$ spin-rotation invariant interactions must have a Larmor mode, the spin magnon, whose energy must tend to $2E_Z$ as $q\to0$. The Larmor mode in Fig.~\ref{fig:FM_excitations} is shown in red. 

\begin{figure}[H]
    \centering \includegraphics[width=0.44\textwidth,height=0.35\textwidth]{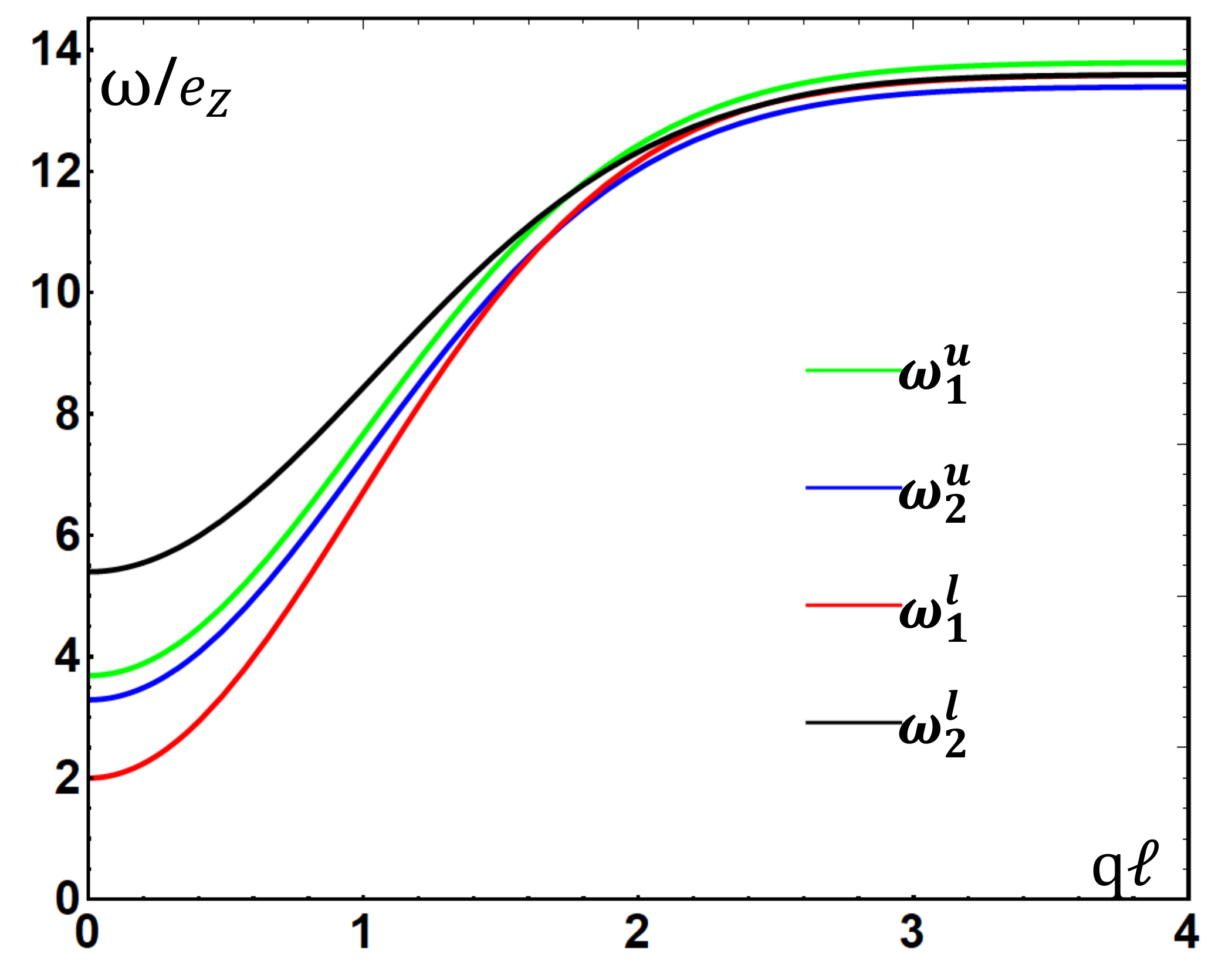}
    \caption{Collective excitations in the \protect\FMcircle\ FM where phase with $E_Z=1e_Z,\ E_V=0.1e_Z$.  The $\omega^u_{1,2}$ are obtained by diagonalizing $\bK^u$ while $\omega^l_{1,2}$ comes from diagonalizing $\bK^l$. The Larmor mode is the one shown in red, and by Larmor's theorem has to have the energy $2E_Z$ at $q=0$.}
    \label{fig:FM_excitations}
\end{figure}

   \item \CDWcircle\ CDW: The eigenvectors and eigenvalues of $H_{MF}$ in the CDW phase are listed below.
\bean
&&U^\text{CDW}=\ 
\begin{blockarray}{ ccccc }
|f_1\rgl &|f_2\rgl &|f_3\rgl &  |f_4\rgl & \\
& & & & &\\
\begin{block}{(cccc)c} 
  1 & 0 & 0 & 0 &  \\
  & & & & &\\
 0 & 1 & 0 & 0 &  \\
 & & & & &\\
 0 & 0 & 1 & 0 & \\
 & & & & & \\
 0 & 0 & 0 & 1 &   \\
\end{block}
\end{blockarray},\\
&&\epsilon_1=u_0-E_Z-E_V-\uzf+2\uzh,\nn
&&\epsilon_2=u_0+E_Z-E_V-\uzf+2\uzh,\nn
&&\epsilon_3=2u_0-E_Z+E_V-2\unf-2\uzh,\nn
&&\epsilon_4=2u_0+E_Z+E_V-2\unf-2\uzh,
\eean 
The transport gap is
\bean\label{cdw_gap}
\Delta^\text{CDW}=u_0-2E_Z+2E_V-2\unf+\uzf-4\uzh,\nn
\eean 
which shows that the gap decreases with $E_Z$ and increases with $E_V$. The collective modes of the CDW phase are shown in Fig.~\ref{fig:CDW_excitations}. Unlike in the case of the FM, there is no Larmor's theorem for the valley magnons, because the interactions are not $SU(2)$ symmetric in the valley space. All collective modes are gapped in the CDW phase because no continuous symmetries are broken. 
\begin{figure}[H]
    \centering \includegraphics[width=0.44\textwidth,height=0.35\textwidth]{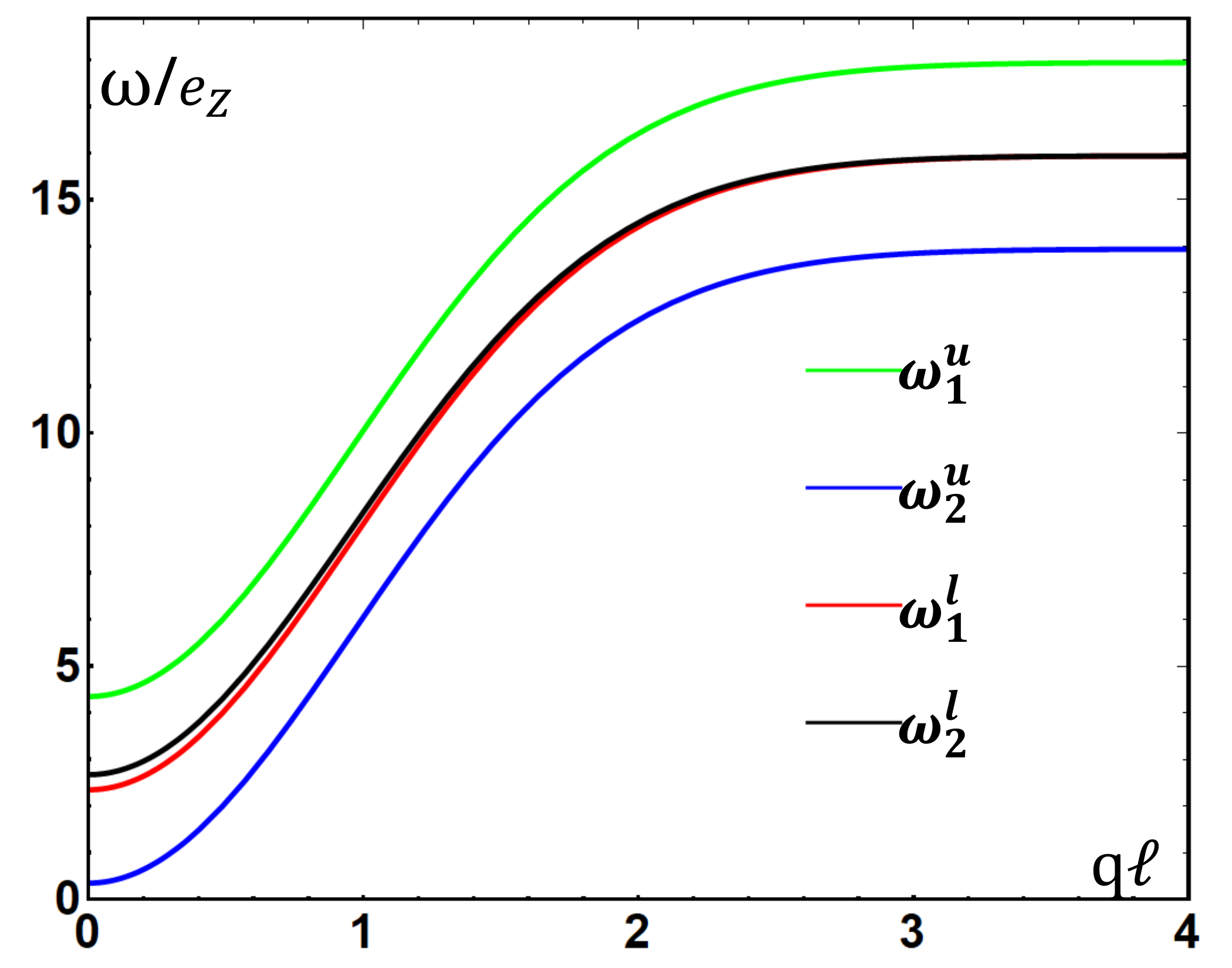}
    \caption{Collective excitations in the  \protect\CDWcircle\ CDW phase with $E_Z=1e_Z,\ E_V=0.1e_Z$.  $\omega^u_{1,2}$ are obtained by diagonalizing $\bK^u$ while $\omega^l_{1,2}$ comes from diagonalizing $\bK^l$. All modes are gapped. Since there is no Larmor's theorem for valley magnons, there is no mode at $q=0$ with an energy of $2E_V$.}
    \label{fig:CDW_excitations}
\end{figure}

  \item \CBOcircle\ CBO: The eigenvectors and eigenvalues of $H_{MF}$ in the CBO phase are listed below.
\bean
&&U^\text{CBO}=\ 
\begin{blockarray}{ ccccc }
|f_1\rgl &|f_2\rgl &|f_3\rgl &  |f_4\rgl & \\
& & & & &\\
\begin{block}{(cccc)c} 
  \cos\frac{\theta_\tau}{2} & 0  & -\sin\frac{\theta_\tau}{2} & 0&  \\
  & & & & &\\
 0 & \cos\frac{\theta_\tau}{2} & 0 & -\sin\frac{\theta_\tau}{2} &   \\
 & & & & &\\
 \sin\frac{\theta_\tau}{2} & 0 & \cos\frac{\theta_\tau}{2} &0 &  \\
 & & & & & \\
 0 & \sin\frac{\theta_\tau}{2} & 0 & \cos\frac{\theta_\tau}{2} &    \\
\end{block}
\end{blockarray},\\
&&\epsilon_1=u_0-E_Z-\unf+2\unh,\nn
&&\epsilon_2=u_0+E_Z-\unf+2\unh,\nn
&&\epsilon_3=2u_0-E_Z-\unf+2\unh-\uzf,\nn
&&\epsilon_4=2u_0+E_Z-\unf+2\unh-\uzf,
\label{eq:UCBO}\eean 
The angle appearing in $U^\text{CBO}$ is defined in Eq.~\eqref{eq:angle_CBO}.
The transport gap in the CBO phase is 
\bean\label{cbo_gap}
\Delta^\text{CBO}=u_0-2E_Z-\uzf.
\eean 
This shows that $\Delta^\text{CBO}$ decreases with $E_Z$ but is independent of $E_V$. The collective modes of the CBO phase are shown in Fig.~\ref{fig:CBO_excitations}. Now the $U(1)_v$ valley symmetry is broken spontaneously, and therefore there is a gapless Goldstone mode, shown in green. However, we note that the $U(1)_v$ symmetry applies only at the four-fermion level. One expects this symmetry to break down to  $Z_3$ when 3-body interactions are included. Thus, in a real graphene sample, one does not expect gapless Goldstone modes in the CBO phase. 

\begin{figure}[H]
    \centering \includegraphics[width=0.44\textwidth,height=0.35\textwidth]{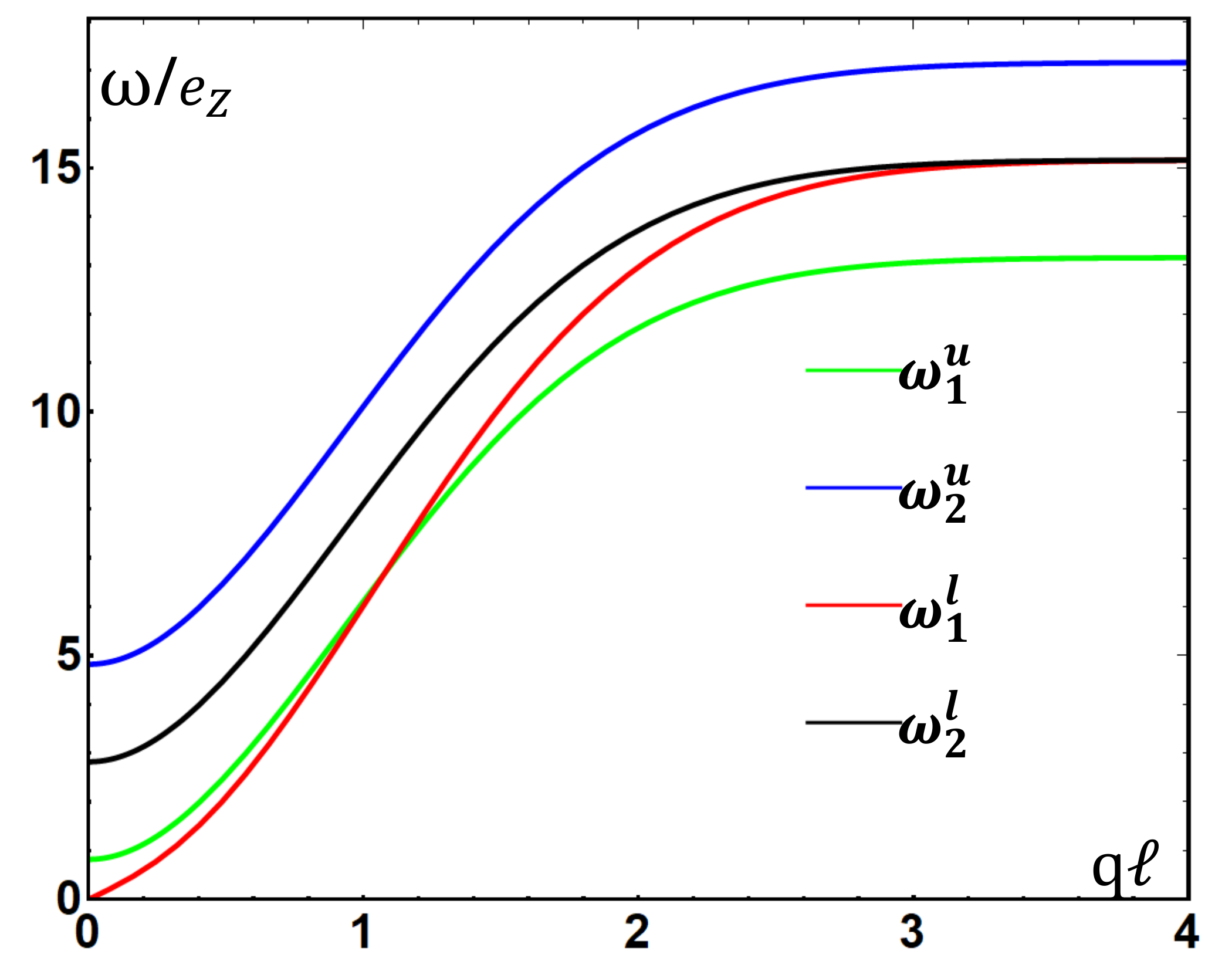}
    \caption{Collective excitations of the \protect\CBOcircle\ CBO phase with $E_Z=1e_Z,\ E_V=0.1e_Z$.   One gapless Goldstone mode is observed due to the spontaneous breaking of the $U(1)_v$ symmetry. This mode is expected to become gapped when 3-body interactions are included. There is no spin polarization, and thus no Larmor mode.}
    \label{fig:CBO_excitations}
\end{figure}

  \item \CAFcircle\ CAFM: The eigenvectors and eigenvalues of $H_{MF}$ in the CAFM state are listed below.
\bean
&&U^\text{CAFM}=\ 
\begin{blockarray}{ ccccc }
|f_1\rgl &|f_2\rgl &|f_3\rgl &  |f_4\rgl & \\
& & & & &\\
\begin{block}{(cccc)c} 
  \cos\frac{\theta_s}{2} & 0 & -\sin\frac{\theta_s}{2} & 0 &  \\
  & & & & &\\
 \sin\frac{\theta_s}{2} & 0 & \cos\frac{\theta_s}{2} & 0 &  \\
 & & & & &\\
 0 & \cos\frac{\theta_s}{2} & 0 & \sin\frac{\theta_s}{2}  & \\
 & & & & & \\
 0 & -\sin\frac{\theta_s}{2} & 0 & \cos\frac{\theta_s}{2} &   \\
\end{block}
\end{blockarray},\\
&&\epsilon_1=u_0-E_V-\uzf,\nn
&&\epsilon_2=u_0+E_V-\uzf,\nn
&&\epsilon_3=2u_0-E_V-2\unf,\nn
&&\epsilon_4=2u_0+E_V-2\unf,
\eean 
where the angle appearing in $U^\text{CAFM}$ is defined in Eq~\eqref{eq:angle_CAFM}. 
The transport gap is
\bean\label{caf_gap}
\Delta^\text{CAFM}=u_0-2E_V-2\unf+\uzf,
\eean 
which  decreases with $E_V$ but is independent of $E_Z$.\\
The collective modes of the CAFM phase are shown in Fig.~\ref{fig:CAF_excitations}. This phase breaks the continuous $U(1)_s$ symmetry of spin rotations around the total field, and therefore has a gapless Goldstone mode, shown in red. In addition, since it has a nonzero spin polarization, it also has a Larmor mode, shown in black. 
\begin{figure}[H]
    \centering \includegraphics[width=0.44\textwidth,height=0.35\textwidth]{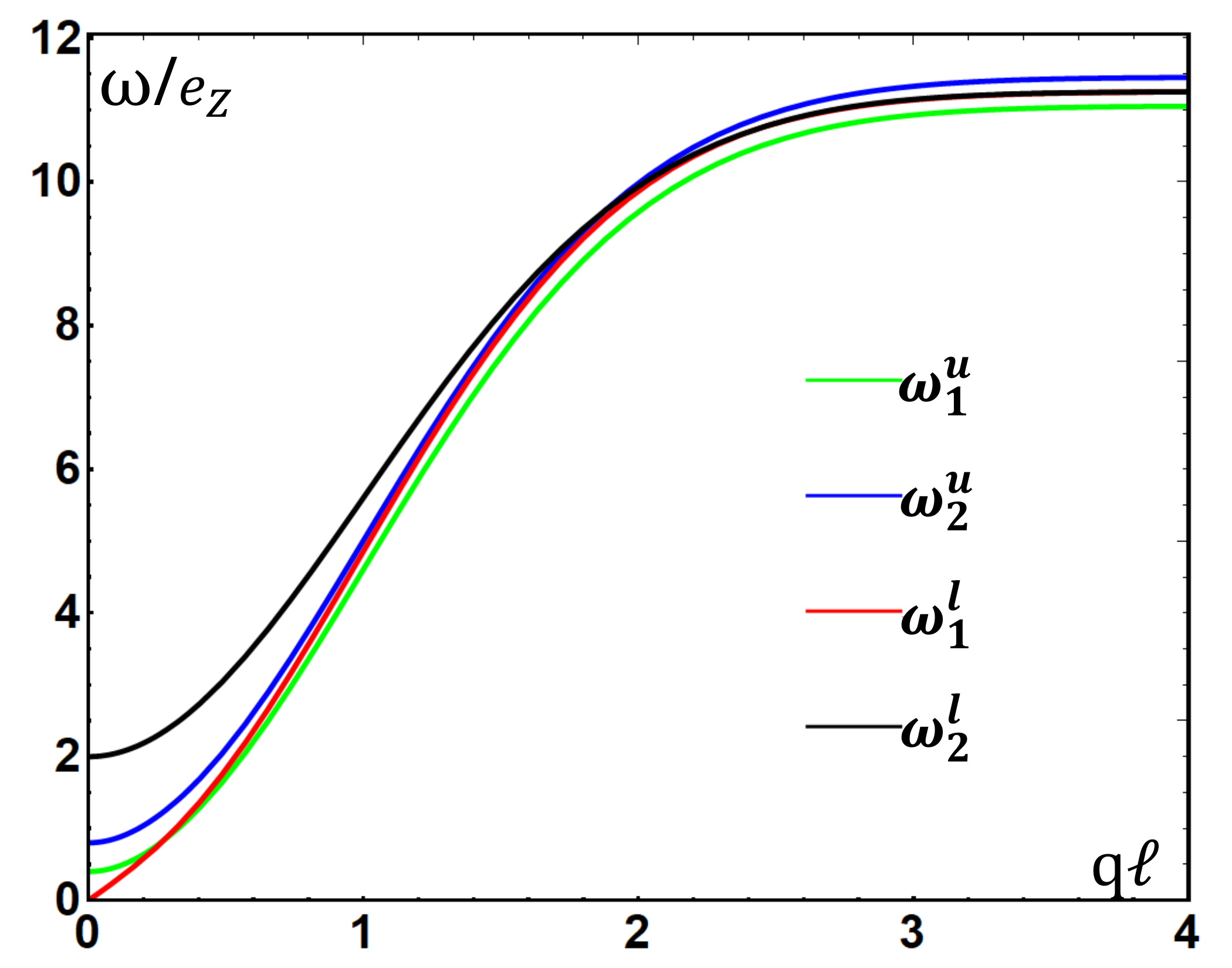}
    \caption{Collective excitations of the  \protect\CAFcircle\ CAFM phase with  $E_Z=1e_Z,\ E_V=0.1e_Z$.  A gapless Goldstone mode (red) corresponding to the spontaneous breaking of the $U(1)_s$ symmetry is observed. The phase has nonzero spin polarization, leading to a Larmor mode (black).}
    \label{fig:CAF_excitations}
\end{figure}
\end{itemize}
In the following subsections, we will present the eigenvectors, transport gaps, and collective excitations of the phases whose occupied spinors cannot be expressed as direct products of spin and valley Bloch spinors.

\subsection{SVE phase}

In the  \SVEcircle\ SVE phase Eq.\eqref{SVE_spinors}, the eigenvectors and eigenvalues  of $H_{MF}[P^{\text{SVE} }]$ are
\bean
&&U^\text{SVE}=\ 
\begin{blockarray}{ ccccc }
|f_1\rgl &|f_2\rgl &|f_3\rgl &  |f_4\rgl & \\
& & & & &\\
\begin{block}{(cccc)c} 
  0 & 1 & 0 & 0 &  \\
  & & & & &\\
 -\sin\frac{\alpha}{2} & 0 & \cos\frac{\alpha}{2} & 0 &  \\
 & & & & &\\
 \cos\frac{\alpha}{2} & 0 & \sin\frac{\alpha}{2} & 0 & \\
 & & & & & \\
 0 & 0 & 0 & -1 &   \\
\end{block}
\end{blockarray},\\
&&\epsilon_1=u_0-u_{\perp,F}-\frac{W}{\uzf-\uzh},\nn
&&\epsilon_2=u_0-u_{\perp,F},\nn
&&\epsilon_3=2u_0-u_{\perp,F}-u_{z,F},\nn
&&\epsilon_4=2u_0-u_{\perp,F}-u_{z,F}+\frac{W}{\uzf-\uzh},\nn
\eean
where the dependence  on the 1-body couplings $E_Z,~E_V$ is contained in $W$,
\bean
W&=&E_V(\unf+\uzf)\nn
&&-(E_Z+\unf+\uzf)(\unf-\uzf+2\uzh).
\eean 
The angle appearing in $U^\text{SVE}$ is defined in Eq.~\eqref{eq:angle_SVE}. 
The transport gap in the SVE phase, 
\bean
\Delta^\text{SVE}=u_0-\uzf,
\eean 
is independent of both $E_Z$ and $E_V$.\\
\begin{figure}[H]
    \centering \includegraphics[width=0.44\textwidth,height=0.35\textwidth]{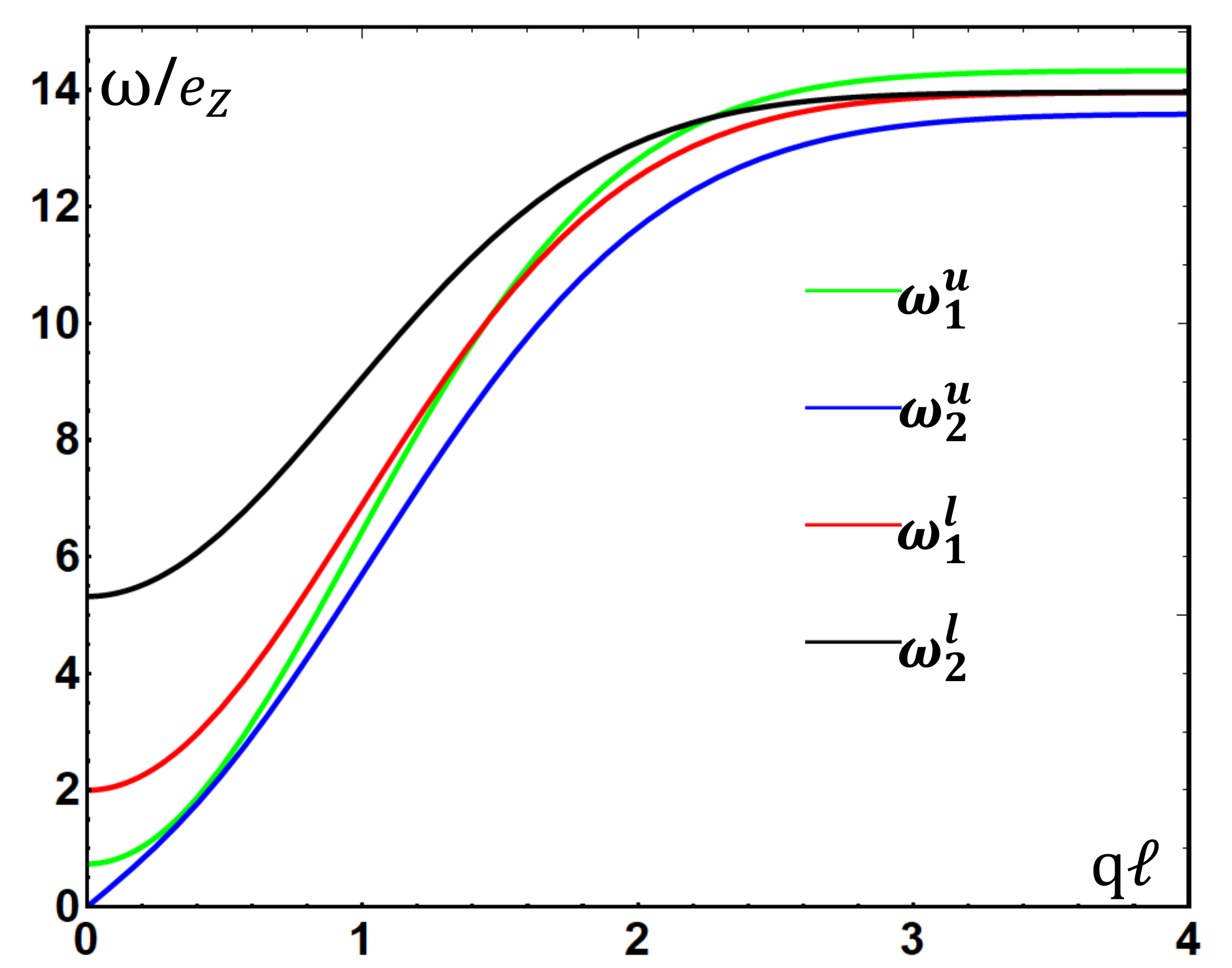}
    \caption{Collective modes of the \protect\SVEcircle\ SVE phase with $E_Z=1e_Z,\ E_V=0.1e_Z$. Since the state breaks a particular linear combination of the $U(1)_s\otimes U(1)_v$ symmetry, only one Goldstone mode exists (black). Furthermore, since the state has nonzero spin polarization, there is a Larmor mode (red).}
    \label{SVE_excitations}
\end{figure}
The dispersion relations of the collective modes are shown in Fig.~\ref{SVE_excitations}. This state breaks a particular $U(1)$ subgroup (generated by $\tau_z-\sigma_z$) of the $U(1)_s\otimes U(1)_v$ symmetry of the Hamiltonian. It thus has only a single Goldstone mode, shown in blue. In addition, it has a nonzero spin polarization, leading to a Larmor mode, shown in red.

\subsection{ SVEX}
In the \SVExcircle\ SVEX phase Eq.\eqref{E-BO-AFM_spinors},  the eigenvalues and eigenvectors of $H_{MF}[P^{\text{SVEX}}]$ are given by 
\bean
&&U^{\text{SVEX}}=\ 
\begin{blockarray}{ ccccc }
|f_1\rgl &|f_2\rgl &|f_3\rgl &  |f_4\rgl & \\
& & & & &\\
\begin{block}{(cccc)c} 
  0 & \cos\frac{\alpha_2}{2} & 0 & \sin\frac{\alpha_2}{2} &  \\
  & & & & &\\
 \cos\frac{\alpha_1}{2} & 0 & \sin\frac{\alpha_1}{2} & 0 &  \\
 & & & & &\\
 -\sin\frac{\alpha_1}{2} & 0 & \cos\frac{\alpha_1}{2} & 0 & \\
 & & & & & \\
 0 & -\sin\frac{\alpha_2}{2} & 0 & \cos\frac{\alpha_2}{2} &   \\
\end{block}
\end{blockarray},\\
&&\epsilon_1=\epsilon_2=u_0-u_{\perp,F},\nn
&&\epsilon_3=\epsilon_4=2u_0-u_{\perp,F}-u_{z,F}.
\eean
The angles appearing in $U^\text{SVEX}$ are defined in Eq.~\eqref{eq:angle_SVEX}.
There are two pairs of degenerate eigenvalues, which are both independent of $E_Z$ and $E_V$. Therefore, the transport gap $\Delta^\text{SVEX}$ is independent of $E_Z,~E_V$.
\begin{figure}[H]
    \centering \includegraphics[width=0.44\textwidth,height=0.35\textwidth]{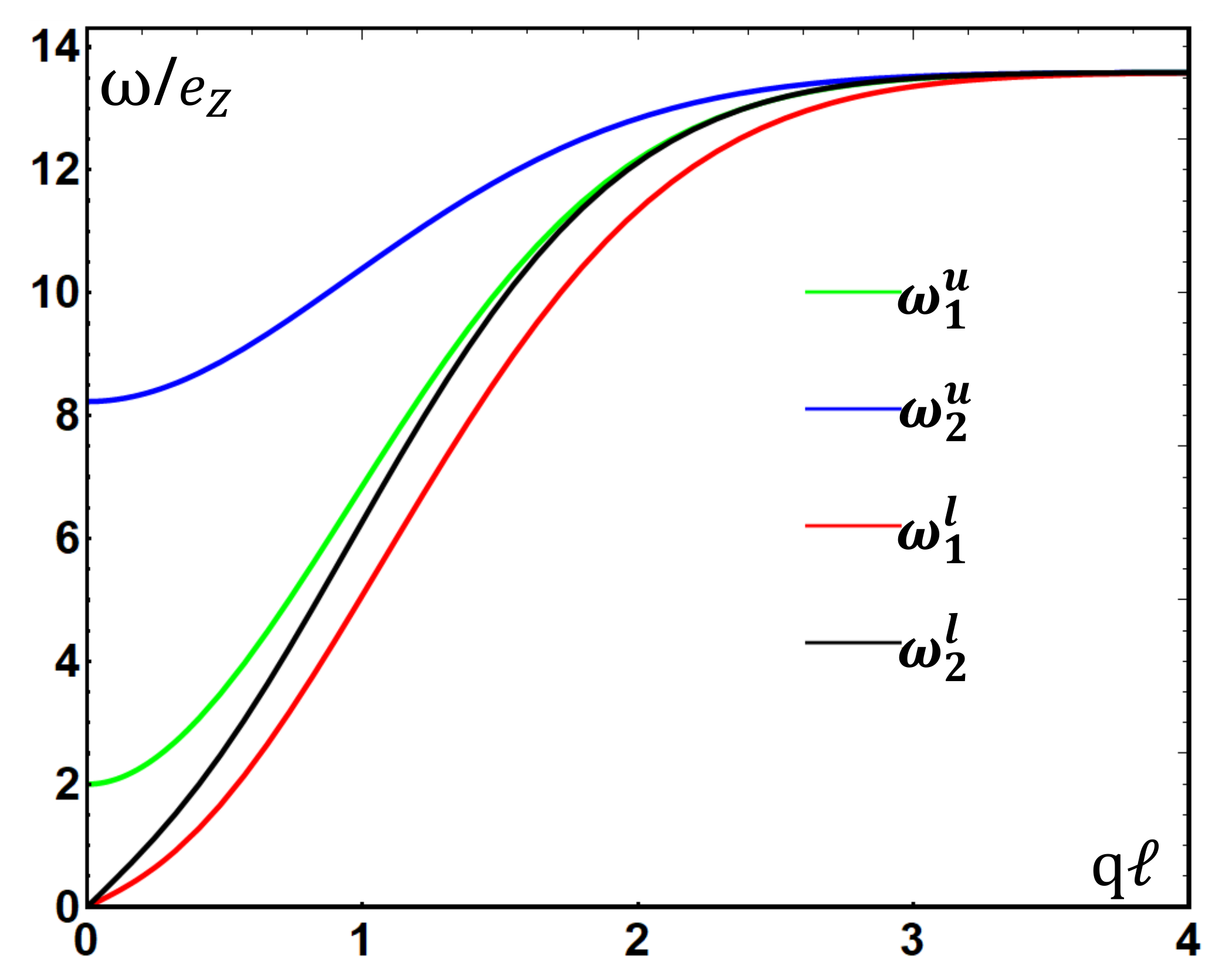}
    \caption{Collective modes of  the \protect\SVExcircle\ SVEX phase with $E_Z=1e_Z,\ E_V=0.1e_Z$. There are two gapless Goldstone modes (red and black) corresponding to the spontaneous breaking of both $U(1)_s$ and $U(1)_v$. Generically, when 3-body interactions are included, the valley Goldstone mode is expected to become gapped. The SVE state also has a nonzero spin polarization, and hence a Larmor mode (green).}
    \label{fig:SVEX_excitations}
\end{figure}

The collective mode dispersions are shown in Fig.~\ref{fig:SVEX_excitations}. The SVEX phase breaks both the spin and valley $U(1)$  symmetries spontaneously, and nominally has two Goldstone modes, shown in red and black. However, do the the inevitable breaking of the valley $U(1)$ to $Z_3$ by 3-body interactions, we expect only one gapless Goldstone mode to be present in real samples. In addition, since the state has a nonzero spin polarization, it has a Larmor mode, shown in green.

\subsection{BO+CAFM phase}
The \protect\BCcircle\ BO+CAFM phase exists at $E_V=0$ and is a simple limit of the COEX phase which occurs when both $E_Z,~E_V\neq0$. \\
Since $\theta_\tau=\frac{\pi}{2}$ in Eq.\eqref{CBO_CAF_spinors} at $E_V=0$, the only nontrivial angles are $\alpha_1,\ \alpha_2$.  The eigenvalues and eigenvectors of $H_{MF}[P^\text{BO+CAFM}]$ are given by
\bean
&&\epsilon_1=u_0-\frac{\uzf}{2}-\unf-E_Z\cos\alpha_1+X_1+R,\nn
&&\epsilon_2=u_0-\frac{\uzf}{2}-\unf+E_Z\cos\alpha_2+X_2+R,\nn
&&\epsilon_3=2u_0-\frac{\uzf}{2}-\unf-E_Z\cos\alpha_2-X_2-R,\nn
&&\epsilon_4=2u_0-\frac{\uzf}{2}-\unf+E_Z\cos\alpha_1-X_1-R,\nn
&&\quad X_{1/2}=\frac{\unh-\unf}{2}\cos 2\alpha_{1/2},\nn
&&\quad R=\frac{\unf}{2}+\frac{\unf-\uzf}{2}\sin\alpha_1\sin\alpha_2\nn
&&\quad\quad\quad+\frac{1}{2}\Big(\unf+2\unh+\uzf\Big)\cos\alpha_1\cos\alpha_2.\\
\quad\nn
&&U=\frac{1}{\sqrt 2}\  
\begin{blockarray}{ ccccc }
|f_1\rgl &|f_2\rgl &|f_3\rgl &  |f_4\rgl & \\
& & & & &\\
\begin{block}{(cccc)c} 
  \cos\frac{\alpha_1}{2} & \sin\frac{\alpha_2}{2}& -\cos\frac{\alpha_2}{2} & \sin\frac{\alpha_1}{2}  &  \\
  & & & & &\\
 -\sin\frac{\alpha_1}{2} & -\cos\frac{\alpha_2}{2} & -\sin\frac{\alpha_2}{2} & \cos\frac{\alpha_1}{2} &   \\
 & & & & &\\
 \cos\frac{\alpha_1}{2} & -\sin\frac{\alpha_2}{2} & \cos\frac{\alpha_2}{2} &\sin\frac{\alpha_1}{2} &  \\
 & & & & & \\
 \sin\frac{\alpha_1}{2} & -\cos\frac{\alpha_2}{2} & -\sin\frac{\alpha_2}{2} & -\cos\frac{\alpha_1}{2}  &   \\
\end{block}
\end{blockarray}.
\eean 
In this case the angles $\alpha_{1,2}$ can be found analytically. The expressions are quite long, and can be found in Appendix C. The collective modes are shown in Fig.~\ref{BO+CAF_excitations}. There are two gapless modes (red and green), which are the Goldstone modes due to the spontaneous symmetry breaking of both $U(1)$ symmetries. As mentioned previously, the valley Goldstone mode is expected to become gapped when 3-body interactions are taken into account. In addition, since the phase has a nonzero spin polarization, there is a Larmor mode (blue).
\begin{figure}[H]
    \centering \includegraphics[width=0.44\textwidth,height=0.35\textwidth]{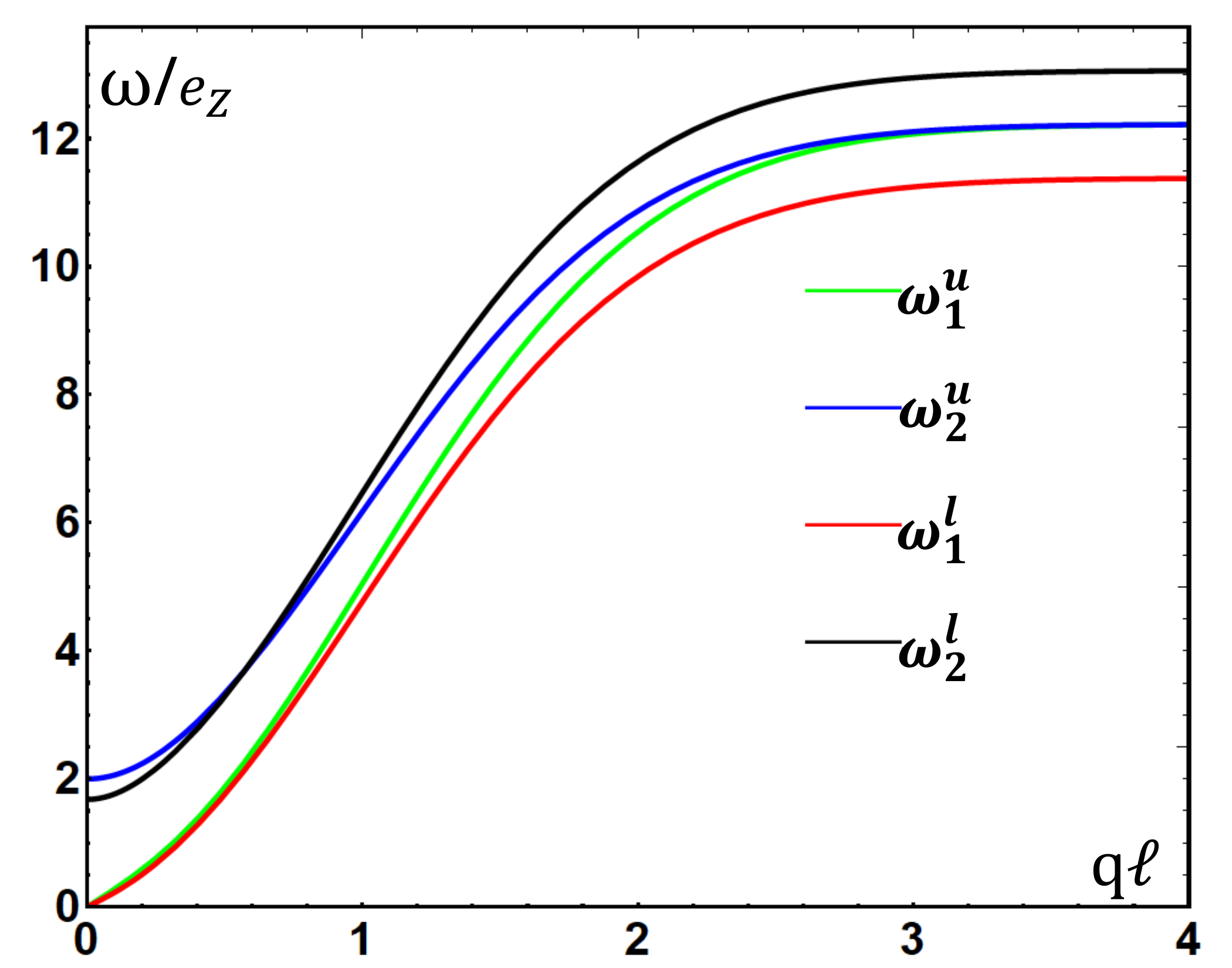}
    \caption{Collective modes of the \protect\BCcircle\  BO+CAFM phase with $E_Z=1e_Z,\ E_V=0$.   Two gapless Goldstone mode are observed (red and green). The valley Goldstone mode is expected to be gapped in a more complete theory. Since the state has nonzero spin polarization, a Larmor mode is also observed (blue). }
    \label{BO+CAF_excitations}
\end{figure}
   
\subsection{COEX phase}

\begin{figure}[H]
    \centering \includegraphics[width=0.485\textwidth,height=0.35\textwidth]{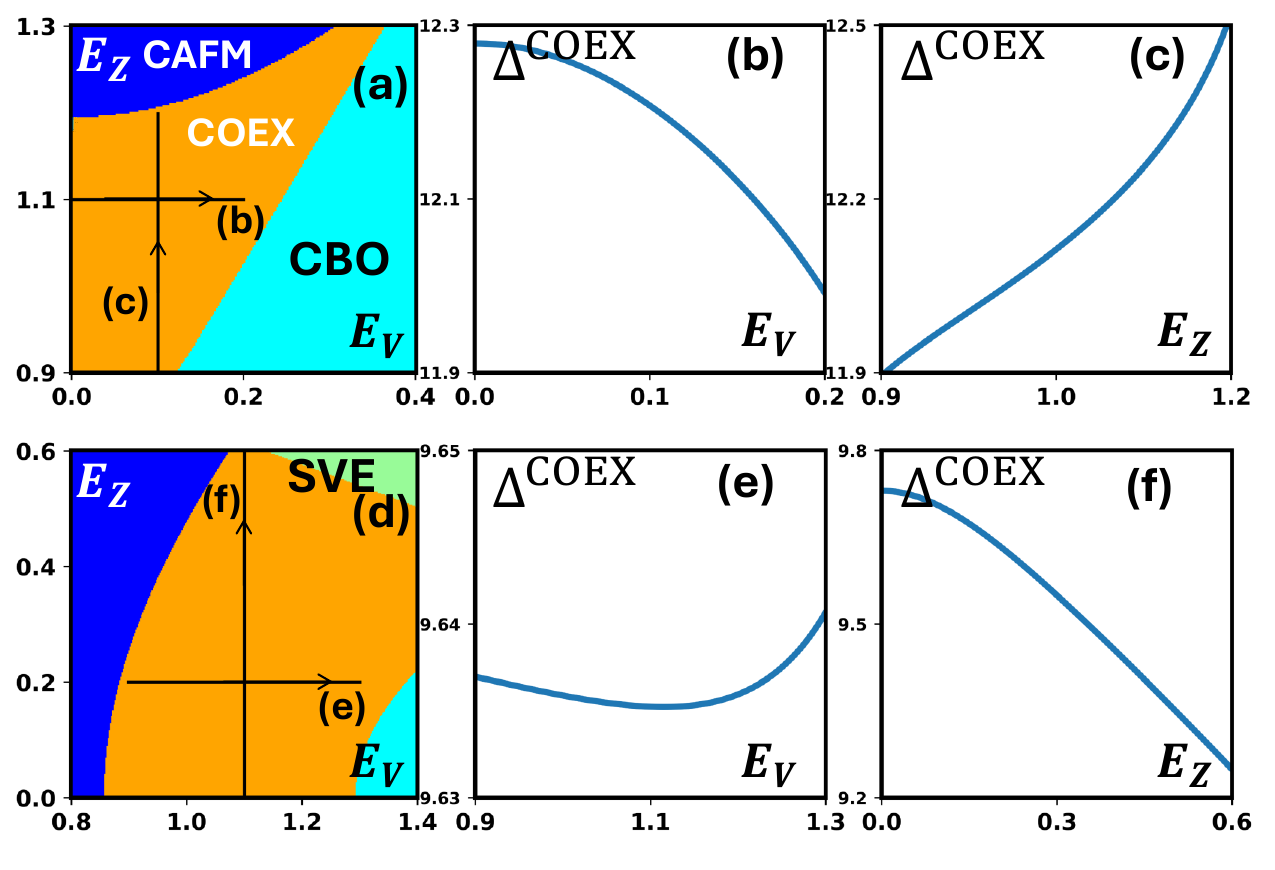}
    \caption{Phase diagrams  and the response of the transport gap to $E_Z,~E_V$ in the COEX phase. The parameters of the anisotropic interactions are fixed at (a) $(g_z,\ g_\perp)=(20,-260)\text{meV}\cdot \text{nm}^{2},\ r_z=r_\perp=-0.2$, (d) $(g_z,\ g_\perp)=(220,-60)\text{meV}\cdot \text{nm}^{2},\ r_z=0.2,\ r_\perp=-0.2$. (b)(e) The transport gap $\Delta^\text{COEX}$ as a function of $E_V$ at fixed $E_Z$.  (c)(f) The transport gap $\Delta^\text{COEX}$ as a function of $E_Z$ at fixed $E_V$. It is seen that the response is not necessarily monotonic. }
    \label{coexisting_gap}
\end{figure}

This is perhaps the most complex phase in the entire phase diagram.  The \protect\CCcircle\ COEX phase  only appears when both $E_Z$ and $E_V$ are nonzero. In addition, numerically this states seems to require all five angles to be described properly. We have been unable to find analytical expressions for the angles or the eigen-energies of the mean-field Hamiltonian in this phase. We therefore have to address this phase numerically.  In Fig. \ref{coexisting_gap}, we present the phase diagrams $(E_Z,\ E_V)$ space for two different values of  $g_z,~g_\perp$. For each such phase diagram, we present the variation of the transport gap with $E_Z$, and with $E_V$ separately. It is seen that the variation of $\Delta^\text{COEX}$ does not show a simple pattern. The gap can either increase or decrease as a function of either $E_Z$ or $E_V$, and can even be non-monotonic.

\begin{figure}[H]
    \centering \includegraphics[width=0.44\textwidth,height=0.35\textwidth]{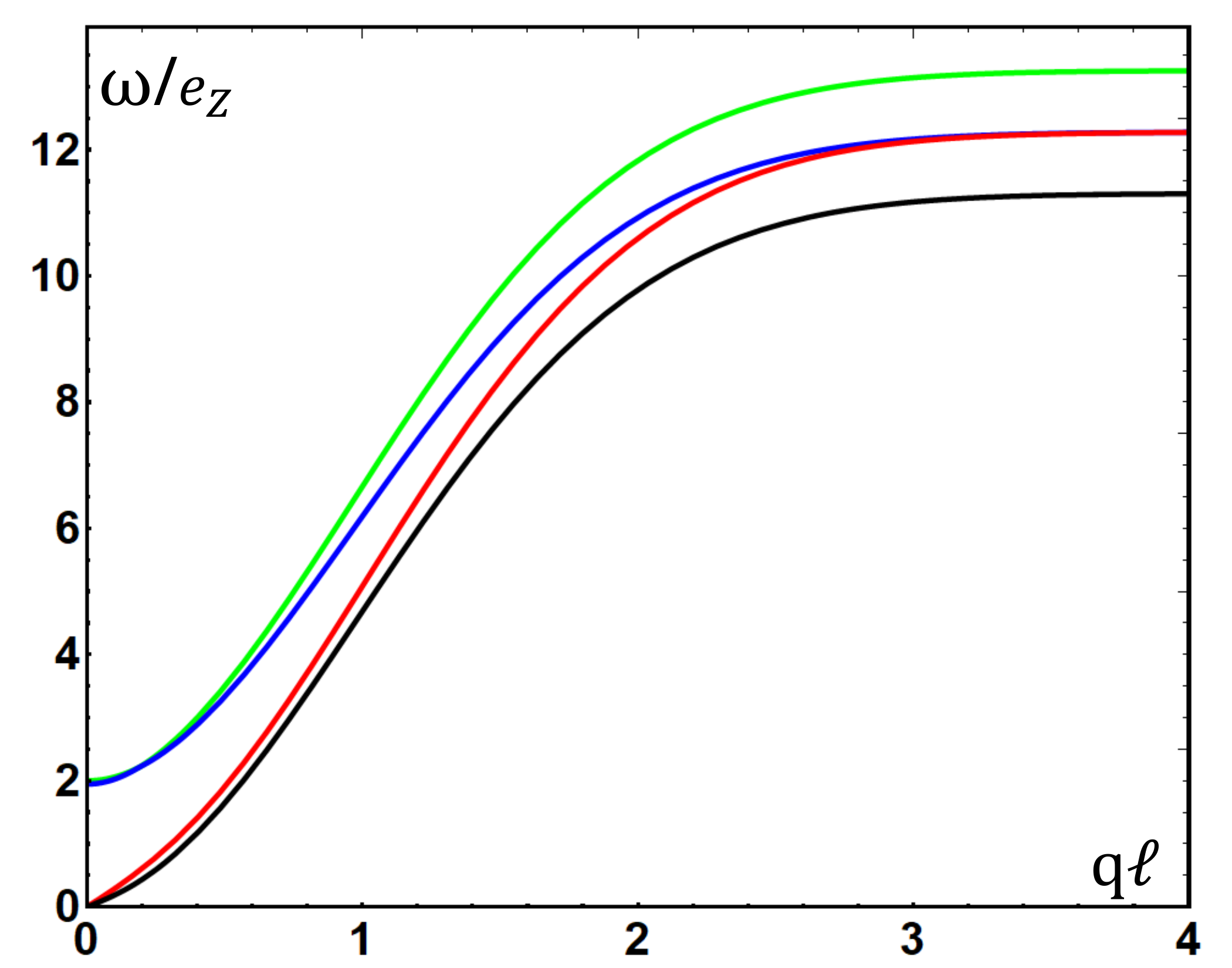}
    \caption{Collective modes in the \protect\CCcircle\ COEX phase at $E_Z=1e_Z,\ E_V=0.1e_Z$.  Two gapless Goldstone modes (red and black) are observed due to the spontaneous symmetry breaking of both the valley and the spin-rotation $U(1)$ symmetries. The valley Goldstone mode will be gapped in a more complete theory. There is also a Larmor mode (blue).  }
    \label{CBO+CAF_excitations}
\end{figure}

Therefore, to identify this phase unambiguously, we need to examine the collective excitations. We have already considered one limit, namely, $E_Z\neq0,~E_V=0$, which is the BO+CAFM phase.   The other extremal case $E_Z=0,~E_V\neq0$, leading to the CBO+AFM phase, which is physically unrealizable, is relegated to Appendix \ref{app: B}. We now present the general case $E_Z,~E_V\neq0$. Fig.~\ref{CBO+CAF_excitations} shows the collective mode dispersions in the COEX phase. Once again, since both the lattice $U(1)$ and the spin-rotation $U(1)$ symmetries are broken spontaneously, there are two Goldstone modes (red and black). Of course, the valley Goldstone mode is expected to become gapped when 3-body interactions reduce the valley symmetry from $U(1)$ to $Z_3$. Since the phase also has a nonzero spin polarization, there is a Larmor mode (blue).  
\section{Conclusions, caveats, and open questions}
\label{sec: discussion_summary}
The $\nu=0$ state of graphene in a strong perpendicular magnetic field has been mysterious for more than a decade. The earliest theories including only Coulomb interactions~\cite{abanin2006:nu0,Brey_Fertig_2006} were shown to be incomplete by two-terminal transport measurements~\cite{young2014:nu0}. A later model by Kharitonov~\cite{kharitonov2012:nu0}, assuming ultra-short-range anisotropic interactions breaking the $SU(4)$ symmetry of the Coulomb interaction down to $SU(2)_\text{spin}\otimes U(1)_\text{valley}$ found four phases, with the real samples conjectured to be in a canted antiferromagnetic phase. This reproduces the phenomenology of the two-terminal measurements in tilted field. The state at purely perpendicular field was then explored by magnon transmission and STM~\cite{Magnon_transport_Yacoby_2018, Spin_transport_Lau2018, Magnon_Transport_Assouline_2021, Magnon_transport_Zhou_2022,li2019:stm, STM_Yazdani2021visualizing, STM_Coissard_2022, Yazdani_FQH_STM2023_1, Yazdani_FQH_STM2023_2}. Furthermore, STM seems to show some form of lattice symmetry breaking most of the time~\cite{li2019:stm, STM_Yazdani2021visualizing, STM_Coissard_2022, Yazdani_FQH_STM2023_1, Yazdani_FQH_STM2023_2}. Since the samples are different in their screening properties - transport samples are doubly encapsulated while STM samples are necessarily open on one face - it is possible that these two classes of samples are in different phases~\cite{Wei_Xu_Sodemann_Huang_LLM_SU4_breaking_MLG_2024,Xu2024}. However, motivated by theoretical renormalization group ideas, and the possibility of phases which break both lattice and magnetic symmetries simultaneously (not found in Kharitonov's model), the ultra-short-range assumption on the anisotropic interactions was relaxed, and it was found that indeed there are several potential phases not present in the USR limit ~\cite{Das_Kaul_Murthy_2020,Das_Kaul_Murthy_2022,Stefanidis_Sodemann2023,Atteia_Goerbig_2021,Hegde_Sodemann2022,Lian_Goerbig_2017}. The order parameters of these phases include bond-order, canted antiferromagnetic order, and also more exotic types of order such as spin-valley entangled order. 

While lattice symmetry breaking, synonymous with valley symmetry breaking in the $n=0$ LL-manifold of graphene, can be directly detected by STM, the other order parameters cannot be imaged directly with present techniques. Thus, there is a pressing need to develop diagnostics for all the potential phases which can be implemented using readily available experimental techniques. 

In this work we provide a complete set of diagnostics that can identify the phase of the system uniquely. The diagnostic criteria consist of two types of measurable quantities: the transport gap, and long-wavelength collective excitations. We assume that $E_Z$ and $E_V$ can be independently varied in situ. This is straightforward for $E_Z$ because applying a parallel field is a well-known technique used in the field. It is also possible to vary $E_V$ in situ in the following way. At perpendicular field, all couplings except $E_V$ are proportional to the strength of the field. Therefore changing the field will alter the ratio of $E_V$ to other couplings, without altering their ratios with each other, effectively serving as a tuning of $E_V$. Clearly, in carrying this out, one has to be mindful that the strength of $B$ affects the screening experienced by the sample. 

We also assume that it is possible to determine whether the system has bulk gapless modes~\cite{Detecting_gapless}. A measurement of the dispersion of the gapless canted antiferromagnon has been carried out in bilayer graphene~\cite{JunZhu_2021}, but we are not aware of a similar measurement in monolayer graphene. There have also been recent attempts to uncover the existence of gapless bulk modes via heat transport, but so far they have not been successful~\cite{Parmentier_2024_Heat_Flow,Anindya_Das_2024_Heat_transport}. 

Last, but not least, the presence of a Larmor mode in the collective mode spectrum signals a nonzero spin polarization, indicating that the phase in question can transport magnons. The presence of a Larmor mode should be detectable in magnon transmission experiments~\cite{Magnon_transport_Yacoby_2018, Spin_transport_Lau2018, Magnon_Transport_Assouline_2021, Magnon_transport_Zhou_2022}. 

Given the assumptions that one can measure the transport gap and its dependence on $E_Z$ and $E_V$ independently, and that one can tell whether there is a neutral gapless bulk mode and a Larmor mode, we have a complete set of diagnostics for determining the phase of the system.  
\begin{widetext} 

\begin{table}[h]
\centering
\begin{tabular}{|p{1.2cm}|p{1.2cm}|p{1.9cm}|p{1.7cm}|p{1.2cm}|p{1.2cm}|p{1.2cm}|p{2.1cm}|p{2.1cm}|p{1.7cm}|}
        \hline
         Phase & \FMcircle\  FM &\AFcircle\ \CAFcircle\ (C)AFM & \BOcircle\ \CBOcircle\ (C)BO &\CDWcircle\ CDW &\SVExcircle\ SVEX &\SVEcircle\ SVE & \BCcircle\ BO+CAFM & \CAcircle\ CBO+AFM & \CCcircle COEX\\
        \hline
    $\quad\frac{d\Delta}{dE_Z}$  & $\quad +$ & $\quad 0$ & $\quad -$  & $\quad -$  & $\quad 0$  & $\quad 0$ & \quad +& Undefined & \quad$\pm$\\
        \hline
    $\quad\frac{d\Delta}{dE_V}$  & $\quad -$ & $\quad -$  & $\quad 0$ & $\quad +$ & $\quad 0$ & $\quad 0$ & Undefined& \quad + &  \quad$\pm$\\
        \hline
        $\quad N_G$ & $\quad$0 & $\quad$1 (1) &$\quad$1 (0) & $\quad$0 & $\quad$2 (1) & $\quad$1 & $\quad$2 (1) &\quad 3 (2) &\quad 2 (1)\\
        \hline 
       $\quad N_L$ & $\quad$1 & $\quad$1 & $\quad$0 & $\quad$0 & $\quad$1 & $\quad$1 & $\quad$1 &\quad 0 & \quad 1\\
        \hline 
\end{tabular}
\label{table1}
\caption{This is identical to Table 1, and presents unique diagnostics for all the theoretically possible phases in graphene at $\nu=0$. The first row tells us the response of the transport gap $\Delta$ to increasing $E_Z$. The entry under CBO+AFM is "Undefined" because this phase only exists at $E_Z=0$. The second row tells us the response of $\Delta$ to $E_V$. The entry under BO+CAFM is "Undefined" because this phase only occurs at $E_V=0$. The third row ($N_G$) tells us the number of Goldstone modes. The first number is the nominal number of gapless modes in our model, assuming that $U(1)_\text{valley}$ is an exact symmetry of the Hamiltonian. The number in $()$ is the actual number of Goldstone modes expected when the valley symmetry is reduced to $Z_3$ with the inclusion of 3-body interactions. The last row ($N_L$) tells us whether there is a Larmor mode, indicating a nonzero spin polarization. Each column is unique, showing that all the phases can be distinguished from each other. 
}\label{phase_tables}
\end{table}
\end{widetext}
The results are summarized in Table \ref{phase_tables}, which is identical to Table \ref{table_1} shown earlier, and is reproduced here for the convenience of the reader. The first row tells whether the slope of the transport gap $\Delta$ is positive/negative/zero with respect to $E_Z$, the second tells us the same with respect to $E_V$, and the third and fourth rows tell us the number of Goldstone and Larmor modes. By virtue of the fact that the columns are all different from each other, we can see that each state can be uniquely distinguished from the others by measuring all four quantities. 

Let us consider the cases most likely to be observed in graphene. It is believed that $g_z>0,~g_\perp<0$ and $g_z\approx|g_\perp|$. Let us assume that one can ensure $E_V=0$ by deliberately misaligning the graphene layer and the HBN layer~\cite{Jung2014OriginOB,hbn_jung_2017,Even_Den_Young2018,RibeiroPalau2018TwistableEW,Finney2019TunableCS}. Then the possible phases are BO, BO+CAFM, and CAFM. One can see from the table that the transport gap $\Delta$ decreases with $E_Z$ in the BO phase, increases with $E_Z$ in the BO+CAFM phase, and is independent of $E_Z$ in the CAFM phase. Thus, in this simple case, one can determine the phase simply by measuring the transport gap in tilted field. For further confirmation, one can directly image the bond order in the BO and BO+CAFM phases by STM, and observe the presence of a Goldstone mode in the BO+CAFM and CAFM phases. Although Table 2 shows a Goldstone mode in the BO phase and two in the BO+CAFM phase, this is a consequence of our model only including 2-body interactions, which leads to a valley $U(1)$ symmetry~\cite{alicea2006:gqhe}. Inclusion of more general, 3-body interactions is expected to break this down to $Z_3$, which means the would-be valley Goldstone mode should be gapped. Thus, in the physical region of coupling constants, at $E_V=0$, the combination of the transport gap being independent of $E_Z$ and the presence of a Goldstone mode uniquely distinguishes the CAFM phase. If the transport gap increases with $E_Z$ and there is a Goldstone mode, then the system is definitely in the BO+CAFM phase. 

Under suitable conditions, for large $E_Z,~E_V$ the system may find itself in a spin-valley entangled phase. The two diagnostics of SVE phases are (i) The transport gap is independent of both $E_Z$ and $E_V$. (ii) There is a gapless Goldstone mode.

To conclude, we have presented diagnostics that can be implemented with currently available experimental techniques~\cite{young2014:nu0,Magnon_transport_Yacoby_2018, Spin_transport_Lau2018,Magnon_Transport_Assouline_2021, Magnon_transport_Zhou_2022,JunZhu_2021,Parmentier_2024_Heat_Flow,Anindya_Das_2024_Heat_transport} that uniquely distinguish all the potential phases of graphene at charge neutrality in a strong perpendicular magnetic field. 

The most obvious generalization of our approach is to find signatures that uniquely identify fractional quantum Hall states in monolayer graphene near charge neutrality. Variationally, it is found that there are many more phases even at $\nu=-\frac{1}{3}$ than there are at $\nu=0$~\cite{Sodemann_MacDonald_2014}, especially when the residual interactions are not ultra short range~\cite{Jincheng2024,an2024fractional}. The smallest gap in such systems is the gap of the fractional component, which can be determined from exact diagonalization. It is possible that the behavior of this gap with $E_Z,~E_V$ will provide some diagnostic information about the nature of the phase, but it remains to be seen whether this is enough to uniquely characterize the phase, in conjunction with information about the collective modes.   

It is natural to generalize this procedure to Bernal-stacked bilayer graphene (BLG)~\cite{neto2009:rmp}. In BLG, in addition to spin and valley (which is approximately locked to layer), one also has the orbital degree of freedom, since the $n=0,1$ Landau levels both have zero energy. It is known that even the inclusion of trigonal warping does not break this degeneracy~\cite{Murthy_BLG_2017}.  We can apply this approach to other 2D materials such as in pentalayer graphene~\cite{FQAH_Pentalayer_Graphene_Ju_2024}, or to TMDs such as $MoTe_2$, where fractional anomalous quantum Hall states have been seen at zero field~\cite{FQAH_MoTe2_Xu_2023a, FQAH_MoTe2_Xu_2023b, FQAH_MoTe2_Mak_Shan_2023, FQAHE_MoTe2_Li_2023}. Of course, in the latter materials the possible two-body couplings may have a different form due to the strong spin-orbit coupling. We plan to study these problems in the near future.

\section{Acknowledgements}
J.A. is supported by the U.S. Department of Energy, Office of Science, Office of Basic Energy Sciences under Award Number DE-SC-0024346. J.A. is also grateful to the University of Kentucky Center for Computational Sciences and Information Technology Services Research Computing for allowing the use of the Morgan Compute Cluster. G.M. thanks the VAJRA scheme of the Science and Engineering Board, Government of India, for support. We are also delighted to thank Ajit Balram and Udit Khanna for fruitful discussions and feedback.

\bibliography{hall}

\begin{appendices}
\setcounter{figure}{0}
\renewcommand{\thefigure}{A\arabic{figure}}
\section{Phase Diagrams}
\label{app: A}
\begin{figure}[H]
    \centering
  \includegraphics[width=0.48\textwidth,height=0.45\textwidth]{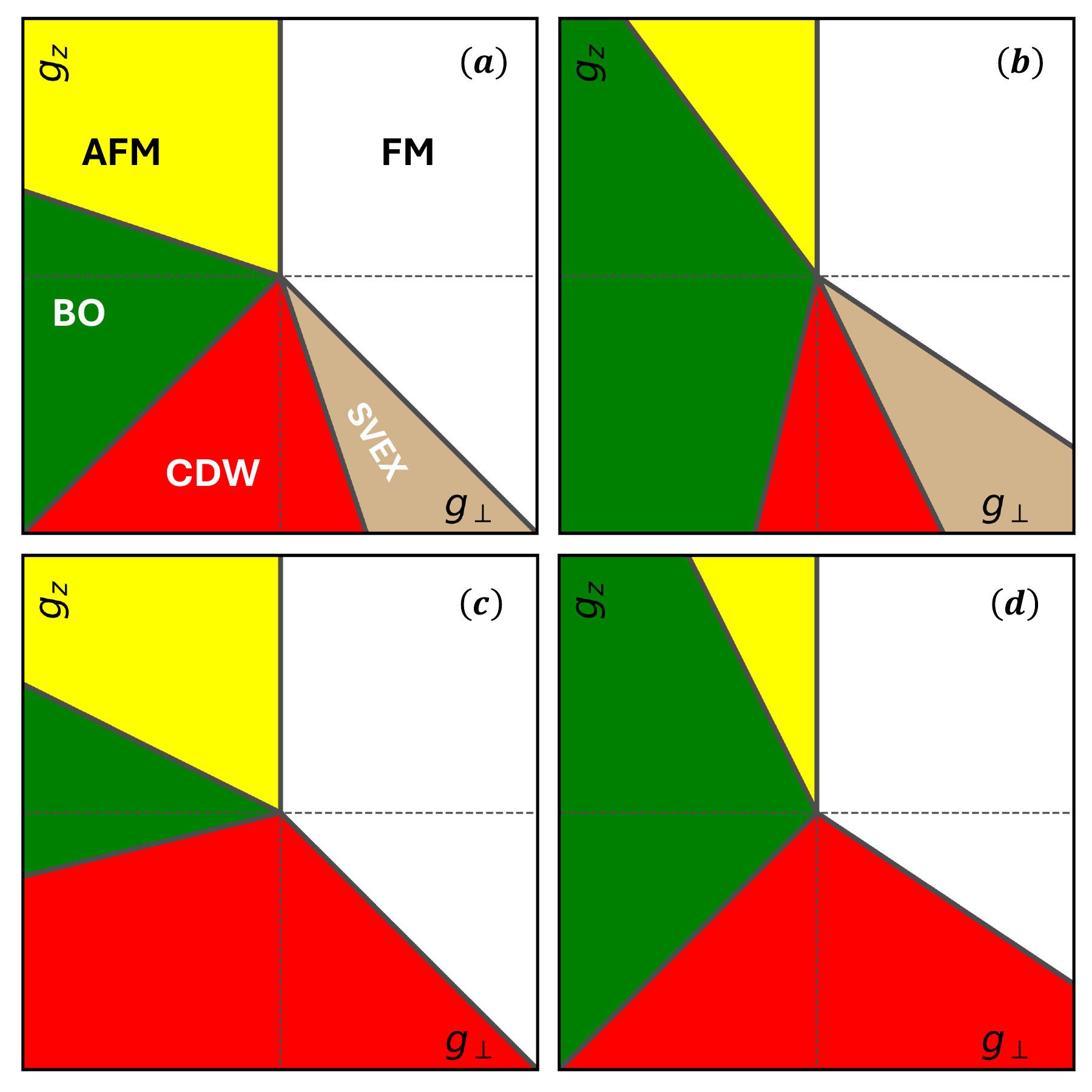}
    \caption{Phase diagrams with $ g_{\perp}\in [-800,800]\text{meV}\cdot \text{nm}^{2}$, $ g_{z}\in [-800,800]\text{meV}\cdot \text{nm}^{2}$ at $E_Z=E_V=0$ and fixed ratios (a) $r_\perp=-0.2,\ r_z=-0.2$, (b) $r_\perp=0.2,\ r_z=-0.2$, (c) $r_\perp=-0.2,\ r_z=0.2$, (d) $r_\perp=0.2,\ r_z=0.2$. All phases will be degenerate at the origin. For negative $r_\perp$, i.e. $\frac{u_{\perp,F}}{u_{\perp,H}}>1$, a SVEX phase will appear.}
    \label{z0v0}
\end{figure}

\begin{figure}[H]
    \centering \includegraphics[width=0.48\textwidth,height=0.45\textwidth]{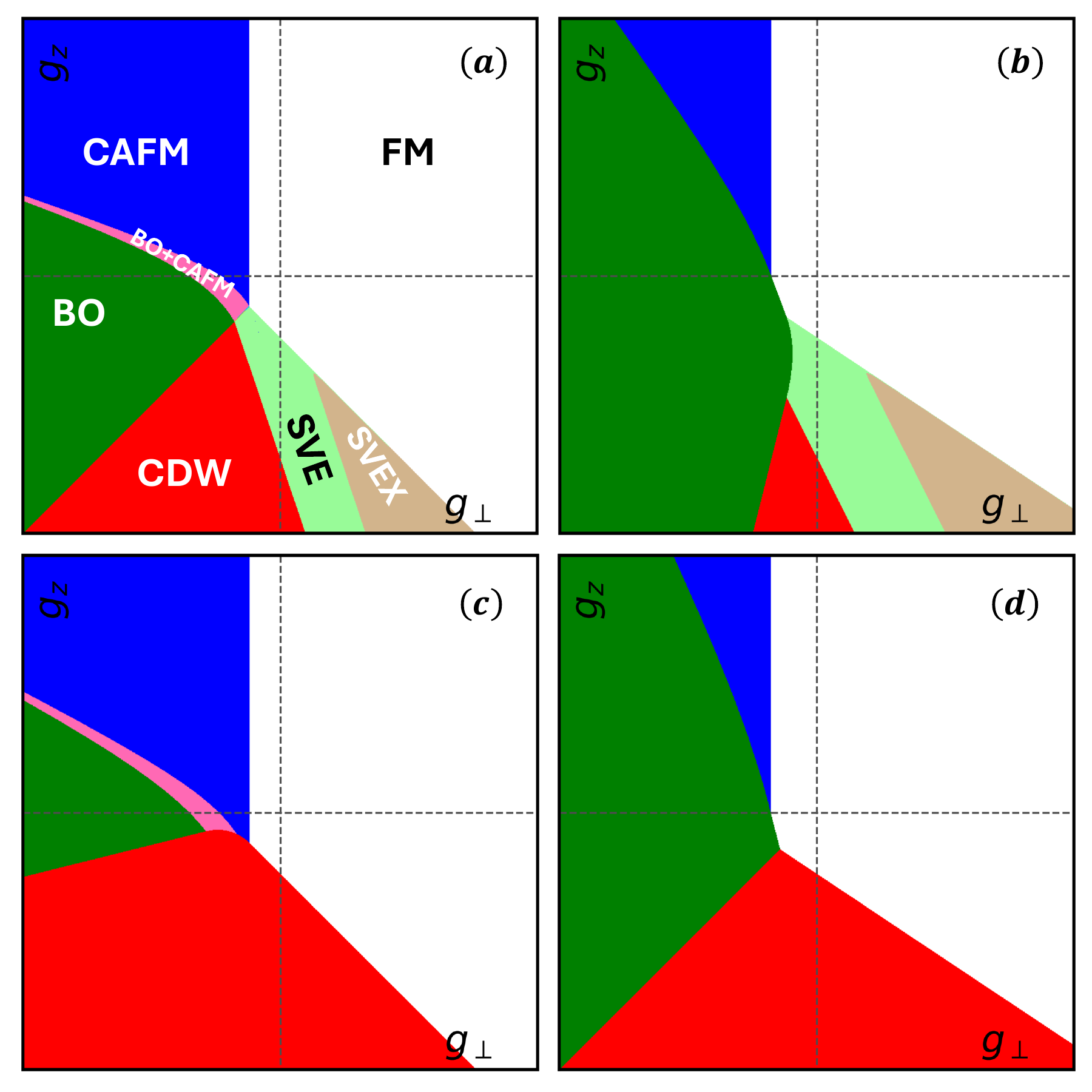}
    \caption{Phase diagrams with $ g_{\perp}\in [-800,800]\text{meV}\cdot \text{nm}^{2}$, $ g_{z}\in [-800,800]\text{meV}\cdot \text{nm}^{2}$ at $E_Z=1e_Z,\ E_V=0$ and fixed ratios (a) $r_\perp=-0.2,\ r_z=-0.2$, (b) $r_\perp=0.2,\ r_z=-0.2$, (c) $r_\perp=-0.2,\ r_z=0.2$, (d) $r_\perp=0.2,\ r_z=0.2$.  }
    \label{z10v00}
\end{figure}

\begin{figure}[H]
    \centering \includegraphics[width=0.48\textwidth,height=0.45\textwidth]{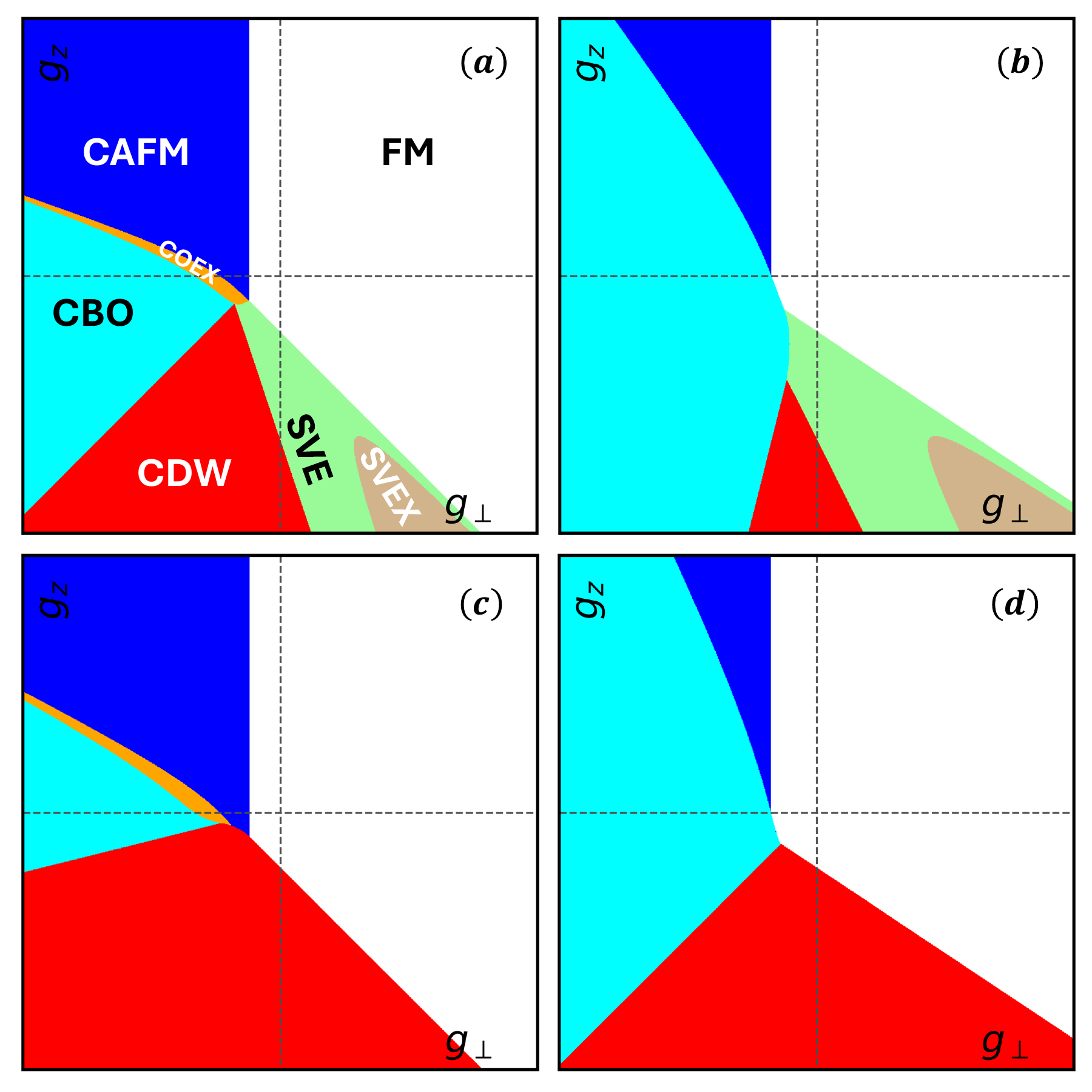}
    \caption{Phase diagrams with $ g_{\perp}\in [-800,800]\text{meV}\cdot \text{nm}^{2}$, $ g_{z}\in [-800,800]\text{meV}\cdot \text{nm}^{2}$ at $E_Z=1e_Z,\ E_V=0.1e_Z$ and fixed ratios (a) $r_\perp=-0.2,\ r_z=-0.2$, (b) $r_\perp=0.2,\ r_z=-0.2$, (c) $r_\perp=-0.2,\ r_z=0.2$, (d) $r_\perp=0.2,\ r_z=0.2$.  }
    \label{z10v01}
\end{figure}

\begin{figure}[H]
    \centering \includegraphics[width=0.48\textwidth,height=0.45\textwidth]{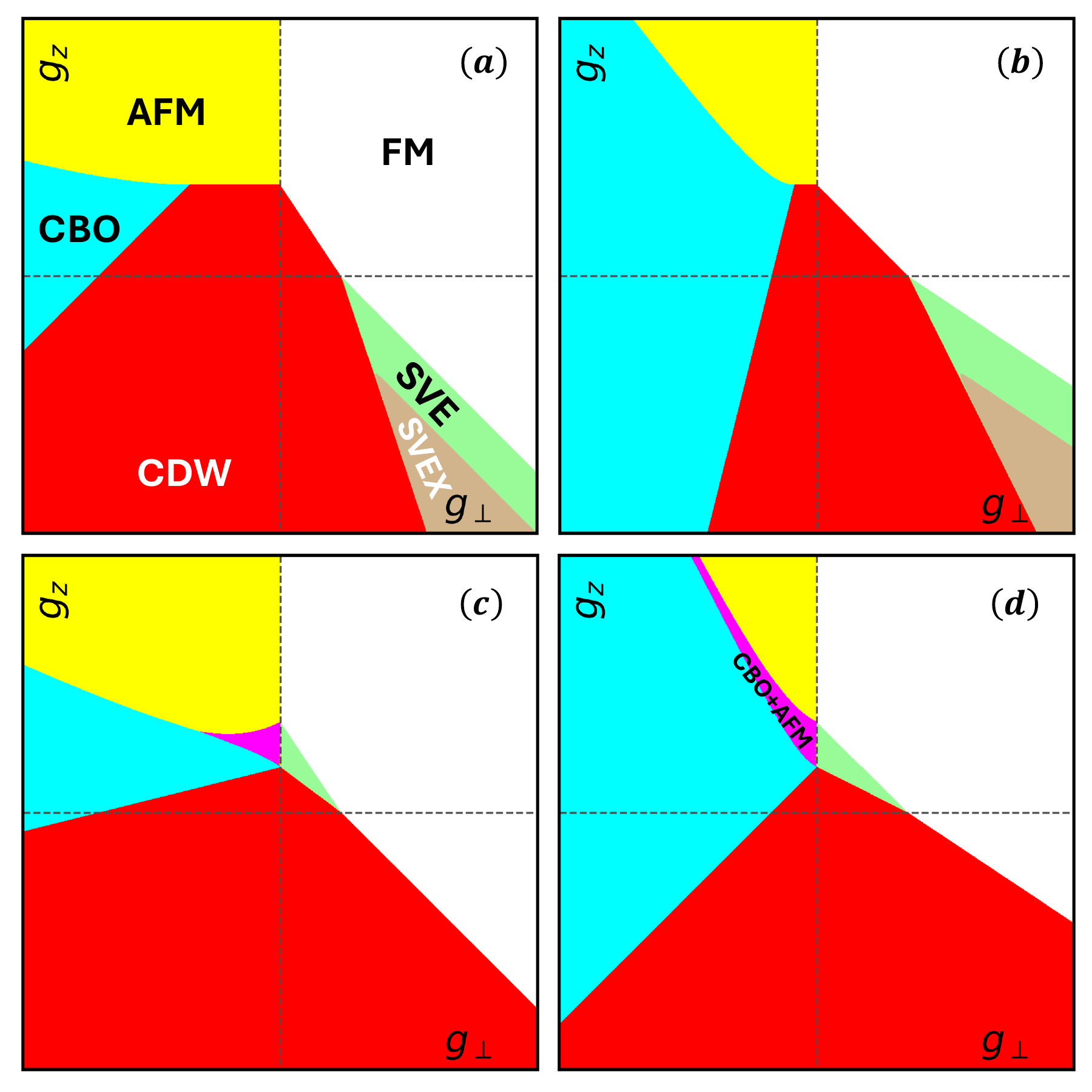}
    \caption{Phase diagrams with $ g_{\perp}\in [-800,800]\text{meV}\cdot \text{nm}^{2}$, $ g_{z}\in [-800,800]\text{meV}\cdot \text{nm}^{2}$ at $E_Z=0,\ E_V=1e_Z$ and fixed ratios (a) $r_\perp=-0.2,\ r_z=-0.2$, (b) $r_\perp=0.2,\ r_z=-0.2$, (c) $r_\perp=-0.2,\ r_z=0.2$, (d) $r_\perp=0.2,\ r_z=0.2$.  }
    \label{z00v10}
\end{figure}

\begin{figure}[H]
    \centering \includegraphics[width=0.48\textwidth,height=0.45\textwidth]{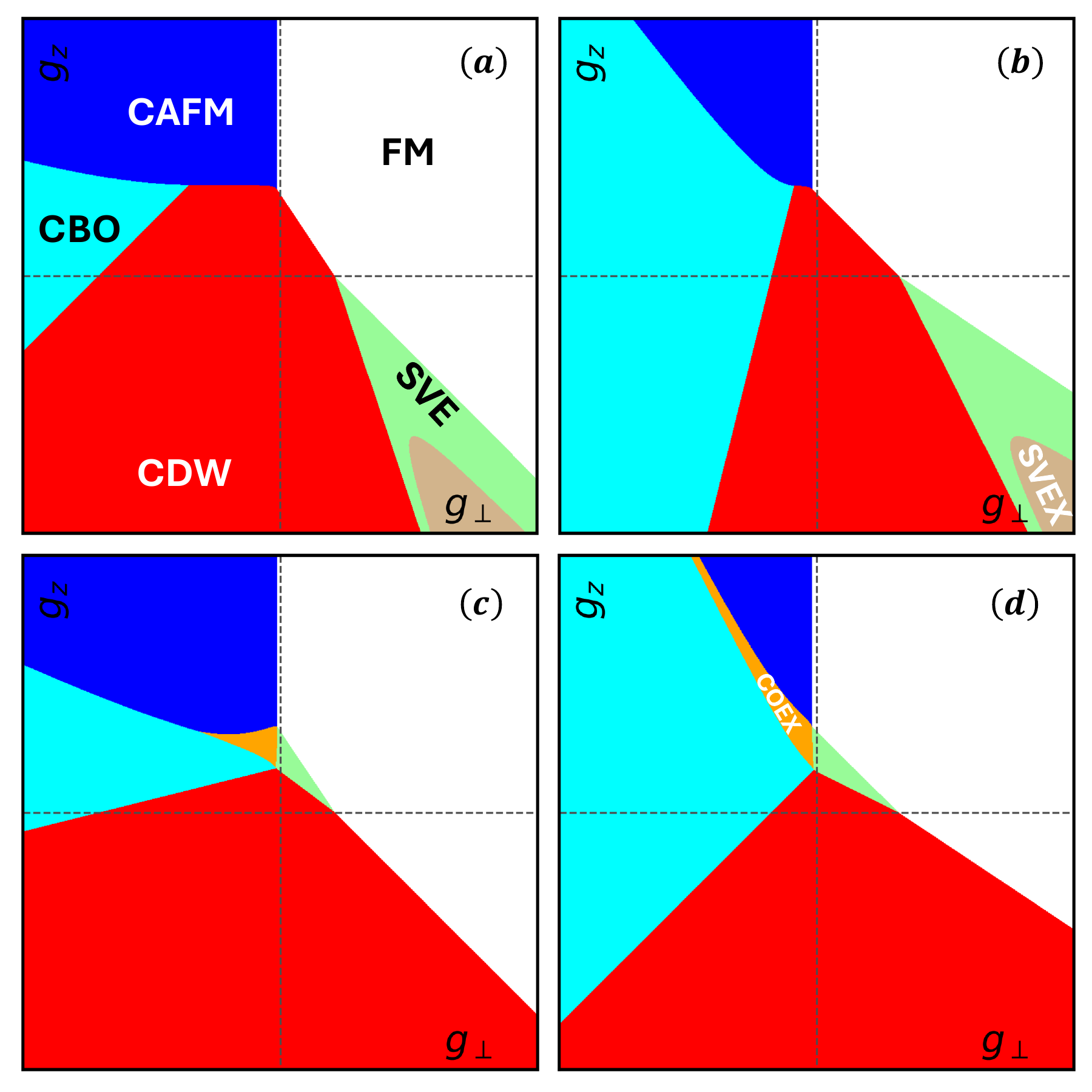}
    \caption{Phase diagrams with $ g_{\perp}\in [-800,800]\text{meV}\cdot \text{nm}^{2}$, $ g_{z}\in [-800,800]\text{meV}\cdot \text{nm}^{2}$ at $E_Z=0.1e_Z,\ E_V=1e_Z$ and fixed ratios (a) $r_\perp=-0.2,\ r_z=-0.2$, (b) $r_\perp=0.2,\ r_z=-0.2$, (c) $r_\perp=-0.2,\ r_z=0.2$, (d) $r_\perp=0.2,\ r_z=0.2$.  }
    \label{z01v10}
\end{figure}

Recall that we characterized the non-USR interactions Eq.\eqref{viq} by the ratio of the Haldane pseudopotentials $r_a=\frac{V_{1a}}{V_{0a}}$. In the main text we focused on both $r_z,~r_\perp>0$ or $r_z,~r_\perp<0$. In this appendix we will present the phase diagrams for all possible combinations of signs of the two ratios, with and without Zeeman or valley Zeeman energy $E_Z,\ E_V$. \\
First, we present phase diagrams without external fields, i.e. $E_Z=E_V=0$ in Fig.~\ref{z0v0}. We remind the reader that our unit of energy $e_Z$ is the physical value of the Zeeman coupling at $B=10T$. In our theoretical phase diagrams, the Zeeman coupling $E_Z$ we use is a tunable parameter which can have unphysical values such as zero.  In the extremal case $E_V=E_Z=0$, the only possible spin-valley entangled state will be the SVEX phase Eq.\eqref{svex_z0v0}. Next, in Fig.~\ref{z10v00} and Fig.~\ref{z10v01}, we present phase diagrams where the Zeeman field is the dominant external field. In Fig.~\ref{z10v00} $E_V=0.1$ while in Fig.~\ref{z10v01} $E_V=0$. Finally, in the Fig.~\ref{z00v10} and Fig.~\ref{z01v10}, we consider the opposite case where valley Zeeman coupling is dominant. In Fig.~\ref{z00v10} $E_Z=0$, while in Fig.~\ref{z01v10} $E_Z=0.1$.

The following broad features can be observed upon comparing the different phase diagrams. The SVEX phase appears only when $r_\perp<0$. If $E_Z=0$ the CBO+AFM phase only appears for $r_z>0$, whereas the SVE phase appears only for $r_z<0$. As soon as $E_Z>0$, which is always the case physically, the BO+CAFM phase appears for $r_\perp<0$. For $E_V=0$,  the SVE phase differentiates itself from the SVEX phase when $r_z<0$. The most complex phases appear when both $E_Z,~E_V\neq0$. Finally, if $E_V\gg E_Z$, the COEX phase disappears. 

\section{Excitations of phases at $E_Z=0$}
\label{app: B}
In the main text, we studied the physical case of $E_Z\neq0$. It is theoretically interesting to study the unphysical limit $E_Z=0$, because in this case the entire Hamiltonian has a $SU(2)$ spin-rotation symmetry. In this appendix, we will present the TDHF collective excitations for phases that appear at $E_Z=0$. At $E_V=0$ these phases are the AFM phase, the FM phase, the SVEX phase, the CDW phase, and the BO phase. When $E_V\neq0$ the BO phase changes to the CBO phase, and the SVE phase appears in the middle of the SVEX phase. Additionally, the CBO+AFM coexistence phase appears for $r_z>0$. Since the other phases are more generic and have been studied in the main text, we will focus in this appendix on the AFM, the SVEX, and the CBO+AFM phases. 
\subsection*{Collective excitations of the AFM phase}
\begin{figure}[H]
    \centering \includegraphics[width=0.44\textwidth,height=0.35\textwidth]{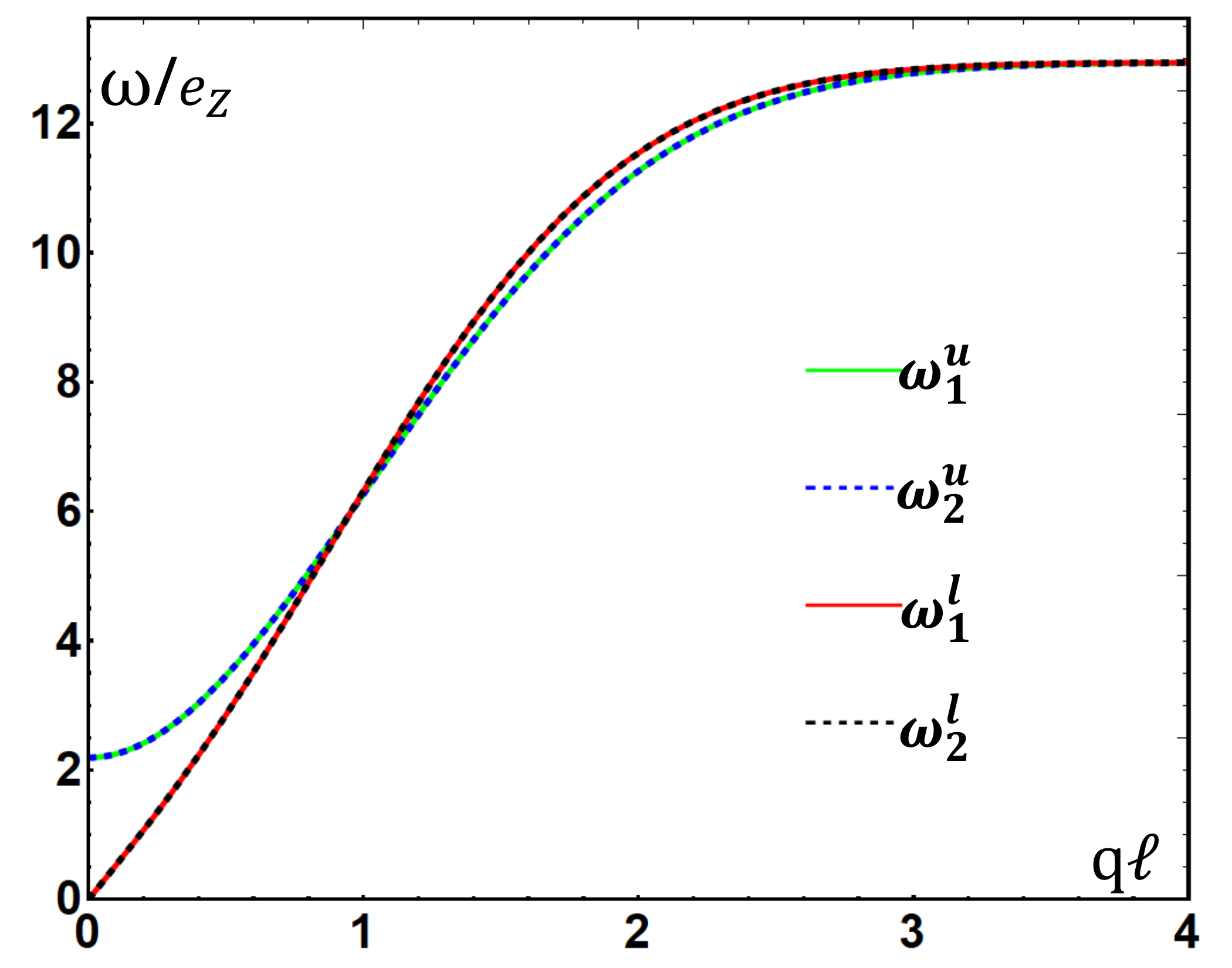}
    \caption{Collective modes of the \protect\AFcircle\ AFM phase with $E_Z=E_V=0$.  The would-be Larmor mode is now a gapless Goldstone mode. There is also the gapless antiferromagnon, bringing the count of the gapless modes to two.}
    \label{AF_excitations}
\end{figure}

Fig.~\ref{AF_excitations} shows the collective mode dispersions in the AFM phase. There are two gapless modes, the Larmor mode which is gapless because $E_Z=0$, and the gapless antiferromagnon. 

\subsection*{Collective excitations of the SVEX phase}
\begin{figure}[H]
    \centering \includegraphics[width=0.44\textwidth,height=0.35\textwidth]{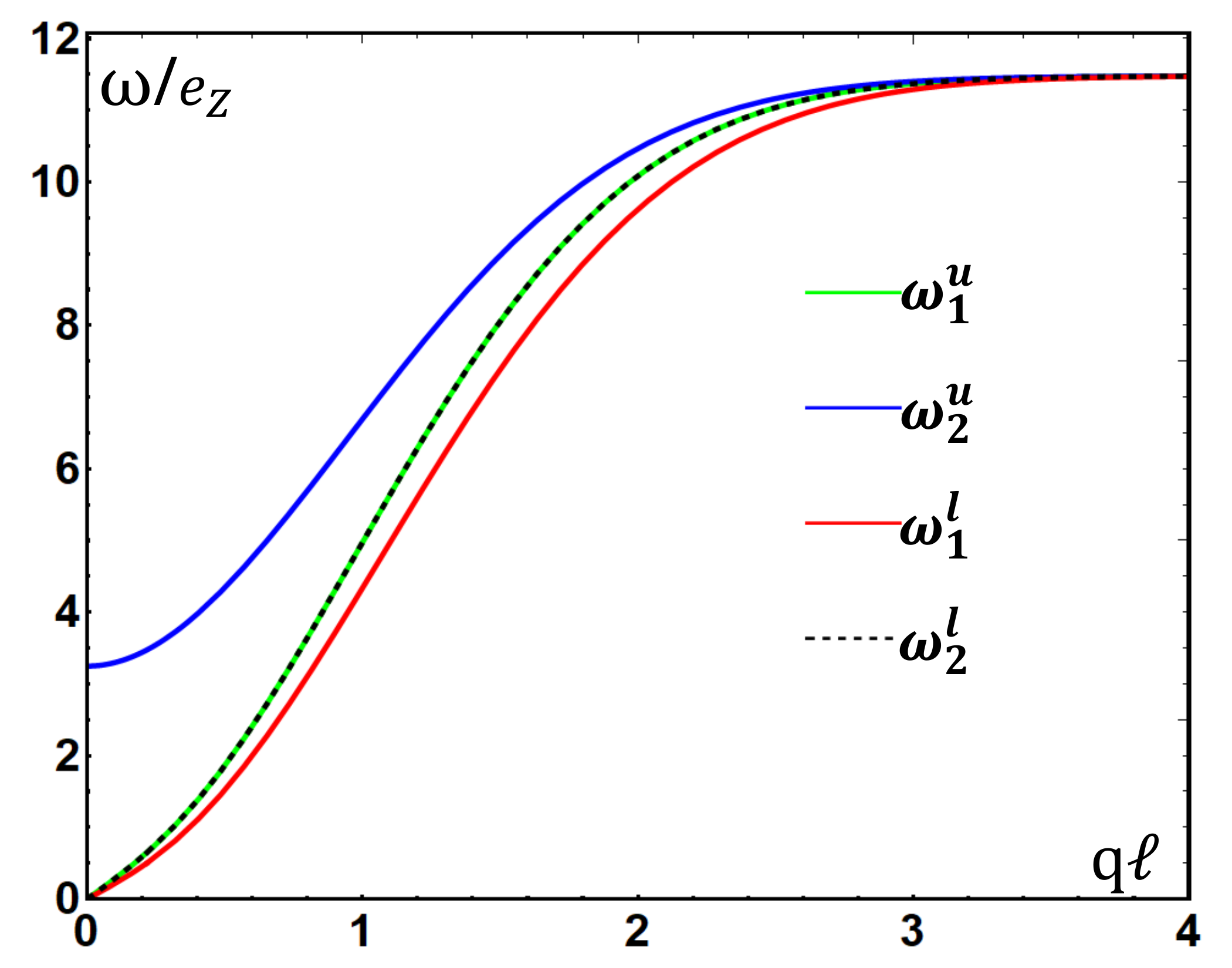}
    \caption{Collective modes of the \protect\SVExcircle\ SVEX phase when $E_Z=E_V=0$.   In comparison with Fig. \ref{fig:SVEX_excitations} in the main text, the Larmor mode at $2E_Z$ now becomes gapless because $E_Z=0$. }
    \label{z0_SVEx_excitations}
\end{figure}

The only spin-valley entangled state that can appear for $E_Z=E_V=0$ is the SVEX phase Eq.\eqref{svex_z0v0}. In fact, since $E_Z=0$, we can make the degeneracy of the ground states due to  $SU(2)_{spin}$ symmetry more obvious by writing the occupied spinors as  
\bean\label{svex_z0v0_app}
\frac{1}{\sqrt 2}\Big(|K,-\bs\rgl-|K^\prime,\bs\rgl\Big),\ \frac{1}{\sqrt 2}\Big(|K,\bs\rgl-|K^\prime,-\bs\rgl\Big),
\eean 
where $\bs$ is an arbitrary unit vector in spin.
This state has no pure spin or valley order. The only nonzero order parameter (for $\bs={\hat e}_z$) is $\lgl\tau_x\sigma_x\rgl$. 
There are two Goldstone modes associated with the ground state spin degeneracy and one Goldstone mode associated with the $U(1)_{v}$ symmetry. As usual, when three-body terms that break the $U(1)_v$ down to $Z_3$ are included, the valley Goldstone mode is expected to become gapped. All this is presented in 
Fig.~\ref{z0_SVEx_excitations}. 

\subsection*{Collective excitations of the CBO+AFM phase}
\begin{figure}[H]
    \centering \includegraphics[width=0.44\textwidth,height=0.35\textwidth]{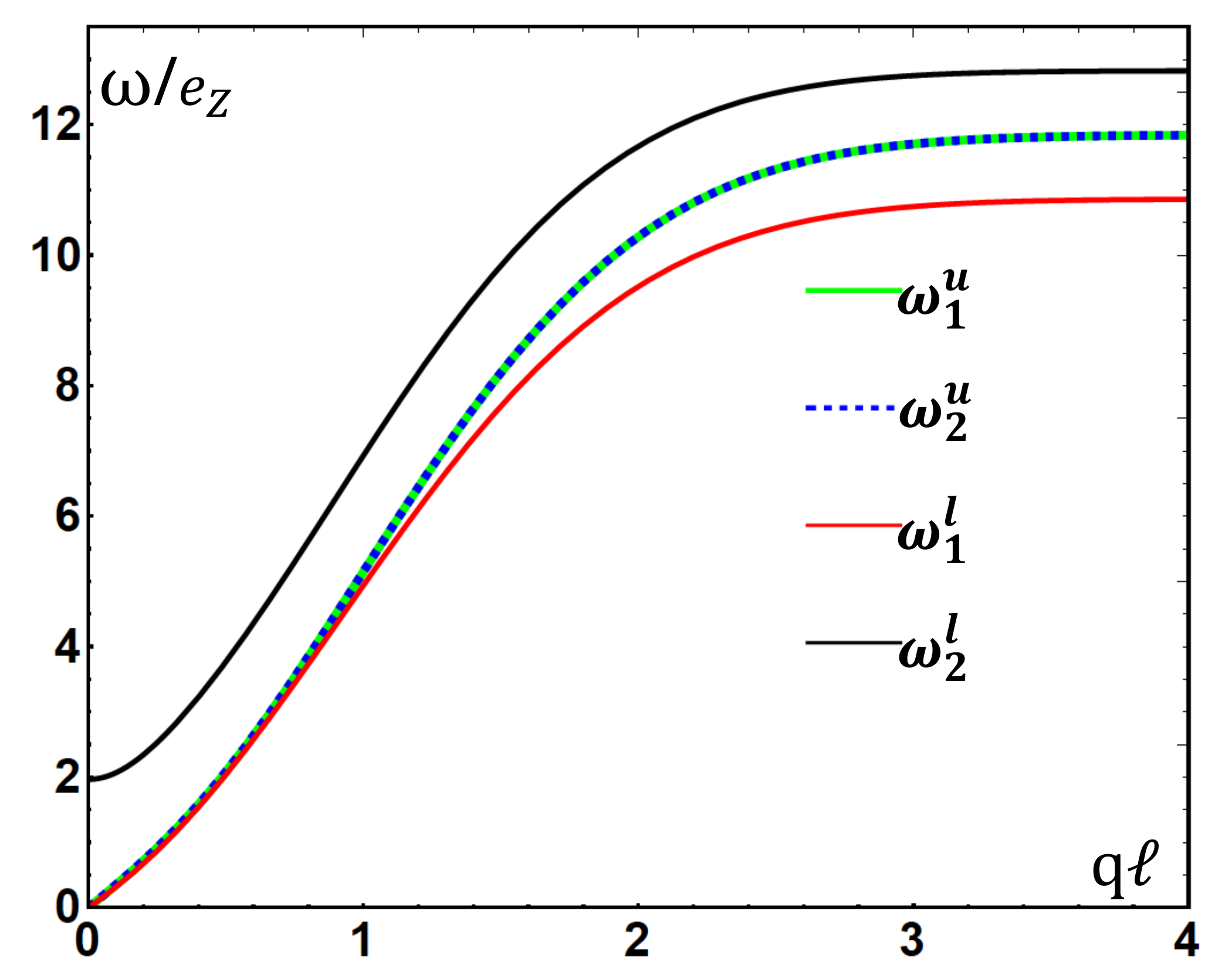}
    \caption{Collective modes of \protect\CAcircle\ CBO+AFM phase where $E_Z=0,\ E_V=1$.  $\omega^u_{1,2}$ are obtained by diagonalizing $\bK^u$ while $\omega^l_{1,2}$ comes from diagonalizing $\bK^l$. 3 gapless Goldstone modes are observed. }
    \label{CBO+AF_excitations}
\end{figure}

In the \protect\CAcircle\ CBO+AFM phase ($\alpha_1=\alpha_2=\alpha$ in Eq.\eqref{CBO_CAF_spinors} at $E_Z=0$), the nontrivial angles are $\alpha,\ \theta_\tau$.  For later convenience, let us define $\alpha^\pm=\alpha\pm\theta_\tau$. In terms of these angles the eigenvalues and eigenvectors of $H_{MF}$ are given by
\bean
&&\epsilon_1=u_0-\unf-\frac{\uzf}{2}+\Lambda+Q^-,\nn
&&\epsilon_2=u_0-\unf-\frac{\uzf}{2}+\Lambda+Q^+,\nn
&&\epsilon_3=2u_0-\unf-\frac{\uzf}{2}-\Lambda-Q^-,\nn
&&\epsilon_4=2u_0-\unf-\frac{\uzf}{2}-\Lambda-Q^+,\nn
&&Q^{\pm}=\frac{\unf-\unh-\uzf+\uzh}{2}\cos2\alpha^{\pm}-E_V\cos\alpha^\pm,\nn
&&\Lambda=\frac{1}{2}\Big(\unf-\unh+\uzh\Big)\nn
&&\quad\quad +\uzh\cos\alpha^+\cos\alpha^--\unh\sin\alpha^+\sin\alpha^-.\\
\quad\nn
&&U=\frac{1}{\sqrt 2}\  
\begin{blockarray}{ ccccc }
|f_1\rgl &|f_2\rgl &|f_3\rgl &  |f_4\rgl & \\
& & & & &\\
\begin{block}{(cccc)c} 
  \cos\frac{\alpha^-}{2} & \cos\frac{\alpha^+}{2} & \sin\frac{\alpha^-}{2} & 
  -\sin\frac{\alpha^+}{2} &  \\
  & & & & &\\
 -\cos\frac{\alpha^-}{2} & \cos\frac{\alpha^+}{2} & -\sin\frac{\alpha^-}{2} & 
 -\sin\frac{\alpha^+}{2} &  \\
 & & & & &\\
 -\sin\frac{\alpha^-}{2} & \sin\frac{\alpha^+}{2} & \cos\frac{\alpha^-}{2} & \cos\frac{\alpha^+}{2} & \\
 & & & & & \\
 \sin\frac{\alpha^-}{2} & \sin\frac{\alpha^+}{2} & -\cos\frac{\alpha^-}{2} & \cos\frac{\alpha^+}{2} &   \\
\end{block}
\end{blockarray}.
\eean 
Compared to Eq.\eqref{CBO_CAF_spinors}, we have made a rotation for both filled $|f_1\rgl,\ |f_2\rgl$ and empty $|f_3\rgl,\ |f_4\rgl$, i.e.
\bean\label{rotation}
&&\begin{pmatrix}
   |f_1\rgl \\
    |f_2\rgl\\
\end{pmatrix}\ \rightarrow\ 
\frac{1}{\sqrt 2}
\begin{pmatrix}
   1 &1\\
  -1 &1\\
\end{pmatrix}
\begin{pmatrix}
   |f_1\rgl \\
    |f_2\rgl\\
\end{pmatrix},\nn
&&\begin{pmatrix}
   |f_3\rgl \\
    |f_4\rgl\\
\end{pmatrix}\ \rightarrow\ 
\frac{1}{\sqrt 2}
\begin{pmatrix}
   1 &1\\
  -1 &1\\
\end{pmatrix}
\begin{pmatrix}
   |f_3\rgl \\
    |f_4\rgl\\
\end{pmatrix},
\eean 
where we have taken $\eta_1=\eta_2=\frac{\pi}{2}$ in Eq.\eqref{linear_combination} for $E_Z=0$.\\
Fig.~\eqref{CBO+AF_excitations} shows the collective mode dispersions in the CBO+AFM phase. In comparison to Fig.~\ref{CBO+CAF_excitations}, the Larmor mode observed there becomes a gapless Goldstone mode as the Zeeman energy vanishes.

\section{Analytic minimization in the BO+CAFM Phase}
\label{app: C}
In this section, we analytically solve for the angles in the BO+CAFM phase, which is firstly proposed in~\cite{Stefanidis_Sodemann2023}. After putting the spinors \eqref{CBO_CAF_spinors} into the energy functional \eqref{Ehf} and set $\theta_\tau=\frac{\pi}{2}$ for $E_V=0$, we get the expression 
\bean
&&E^\text{BO+CAFM}\nn
&=&E_Z\big(\cos\alpha_2-\cos\alpha_1\big)+\frac{1}{2}\big(\unh-2\unf-\uzf\big)\nn
&&+\frac{\unf+\uzf}{2}\cos\big(\alpha_1+\alpha_2\big)\nn
&&+\frac{\unh+\unf}{2}\cos\big(\alpha_1-\alpha_2\big)\nn
&&+\frac{\unh-\unf}{4}\big(\cos2\alpha_1+\cos2\alpha_2\big).
\eean 
We first perform a coordinate transformation as follows
\bean
&&x_1=\cos\alpha_1,\  y_1=\sin\alpha_1,\nn
&&x_2=\cos\alpha_2,\  y_2=\sin\alpha_2,
\eean
with constraints 
\bean
x_1^2+y_1^2=1,\ x_2^2+y_2^2=1.
\eean 
Next, we have a second transformation 
\bean
z_1=x_1+x_2,\ z_2=x_1-x_2,\nn
w_1=y_1+y_2,\ w_2=y_1-y_2.
\eean 
Then the constrain will become 
\bean
z_1z_2+w_1w_2=0,\ z_1^2+z_2^2+w_1^2+w_2^2=4,
\eean 
one solution of which will be
\bean\label{z1w2}
z_1^2=\bigg(\frac{4}{w_1^2+z_2^2}-1\bigg)w_1^2,\ w_2^2=\bigg(\frac{4}{w_1^2+z_2^2}-1\bigg)z_2^2.\nn
\eean 
Now $E^\text{BO+CAFM}$ will become a function of $w_1$ and $z_2$, 
\bean
&&E^\text{BO+CAFM}\nn
&=&-E_Z z_2-\unf+\frac{2\unh w_1^2}{w_1^2+z_2^2}\nn
&&+\frac{\unf-2\unh-\uzf}{4}w_1^2-\frac{\unf+\uzf}{4}z_2^2.
\eean 
With the above form, we can do the minimization directly, the optimal energy is 
\bean
&&E^\text{BO+CAFM}\nn
&=&\unh-\unf+\frac{E_Z^2+2\unh\big(\unh+\uzf\big)}{2\big(\unf-\unh\big)}\nn
&&+\frac{\sqrt 2E_Z\sqrt{\unh\big(2\unh-\unf+\uzf\big)}}{|\unf-\unh|},
\eean 
which will happen at 
\bean
w_1^2&=&\frac{2\sqrt 2E_Z\unh\big(\unh+\uzf\big)}{\big(\unh-\unf\big)^2\sqrt{\unh\big(2\unh+\uzf-\unf\big)}}\nn
&&-\frac{E_Z^2+2\unh\big(\unf+\uzf\big)}{\big(\unh-\unf\big)^2},\nn
z_2&=&\frac{E_Z}{\unh-\unf}-\frac{\sqrt{2\unh\big(2\unh+\uzf-\unf\big)}}{|\unh-\unf|}.\nn
\eean 
Then combining with Eq.\eqref{z1w2}, we obtain the optimal angles $\alpha_1,\ \alpha_2$ as follows 
\bean
\cos\alpha_{1/2}=\frac{z_1\pm z_2}{2}=\frac{1}{2}\bigg[\sqrt{\Big(\frac{4}{w_1^2+z_2^2}-1\Big)w_1^2}\pm z_2\bigg].
\eean

\end{appendices}

\end{document}